\pdfoutput=1 

\documentclass[
  aps, pre,  twocolumn,  groupedaddress, superscriptaddress 
  showpacs, letterpaper, amsfonts, footinbib, 10pt,  floatfix,
 noeprint]{revtex4-1}

\usepackage{hyperref}
\hypersetup{
    colorlinks=true,
    citecolor=blue,
    linkcolor=black,
    filecolor=cyan,      
    urlcolor=magenta,
    pdfcreator = {\LaTeX\ and \flqq hyperref\frqq},
}

\RequirePackage{graphicx,amsmath,amssymb,bm}
\usepackage{bbm}
\usepackage{float}
\usepackage{framed} 
\usepackage{amsthm}
\usepackage[normalem]{ulem}
 
 \graphicspath{ {./Figs/} } 
 
\newcommand{\bb}{\mathbb}
\newcommand{\tr}{\text{tr}}
\newcommand{\Tr}{\text{Tr}}

\newcommand{\ii}{\text{i}}
\newcommand{\La}{{\cal{L}}}
 
\newcommand{\sn}{\text{sn}}  
\newcommand{\1}{\mathbbm{1}}

\def\be{\begin{equation}}
\def\ee{\end{equation}} 
\def\bsh{\begin{shaded}}
\def\esh{\end{shaded}} 
\def\bpm{\begin{pmatrix}}
\def\epm{\end{pmatrix}}

\begin{document}

\title{Integrable nonunitary quantum circuits}
\author{Lei~Su}
\affiliation{%
  Department of Physics, University of Chicago, Chicago, IL 60637, USA}
\author{Ivar~Martin}
\affiliation{%
  Materials Science Division, Argonne National Laboratory, Argonne, IL 08540, USA}

\begin{abstract} 
We show that the integrable Lindblad superoperators found recently can be used to build integrable nonunitary quantum circuits with two-site gates by demonstrating that the $\check{R}$-matrices are completely positive and trace preserving. Using the bond-site transformation, we obtain the corresponding integrable nonunitary quantum circuits with three-site gates.  When restricted to the diagonals of the density matrix, these quantum models reduce to integrable classical cellular automata.
This approach may pave the way for a systematic construction of integrable nonunitary quantum circuits.
\end{abstract}

\maketitle

\section{Introduction}

Nonequilibruim quantum dynamics has been a focus of attention in theoretical and experimental physics in the recent years. Compared to their equilibrium counterparts, nonequilibrium systems are 
more challenging to describe due to the explicit time-dependence of their evolution. That generally involves coherent superposition of a large number of energy eigenstates, and not only their statistical mixture. 
One key question has to do with the ways an interacting quantum system reaches \cite{deutsch1991quantum, srednicki1994chaos} (or evades \cite{moudgalya2021quantum}) thermodynamic equilibrium. 
Significant progress \cite{Polkovnikov2011,nandkishore2015many,d2016quantum, Abanin2019colloquium, Castro2016emergent, Bertini2016transport, Bertini2021finite, Alba2021genearlized} has been achieved in understanding of the thermalization both in continuous and discrete models, especially in one-dimensional (1D) quantum systems. Some advanced numerical methods have been developed \cite{schollwock2005density, schollwock2011density, Pekker2014hilbert, cirac2021matrix}, yet they remain limited in accessible system sizes and evolution times. A number of analytical results have been obtained, especially in integrable systems  where an extensive number of conserved quantities  strongly constrain the dynamics \cite{baxter2016exactly, vsamaj2013introduction, eckle2019models}. There, the generalized hydrodynamics approach, originally introduced to study inhomogeneous quantum quenches in integrable many-body systems, has been very fruitful \cite{Castro2016emergent, Bertini2016transport, Bertini2021finite, Alba2021genearlized}.   Exact analytical results have  been obtained recently in systems including the Lieb-Liniger model \cite{Granet2021systematic} and the folded XXZ model \cite{pozsgay2021integrable, Zadnik2021folded1, Zadnik2021folded2}.

Typically, integrating time evolution involves discretization of time, with the small time step  ensuring that only small changes in the state of the system during a time step occur.
However, discretizing time, without assuming smallness of the time step, can also be useful for modeling physical systems.
Thus obtained quantum cellular automata (QCA) are locality-preserving maps that act in discrete space and time \cite{farrelly2020review, piroli2020quantum}, and can be regarded as a special kind of tensor network \cite{piroli2020quantum, cirac2021matrix}. The focus of this paper are QCA that are constant-depth quantum circuits, which consist of layers of gates acting on a lattice of qubits \cite{farrelly2020review, cirac2021matrix}. Several exact results have already been obtained in such quantum circuits, including random quantum circuits \cite{nahum2017quantum, nahum2018operator, vonkeyserlingk2018operator}, dual unitary circuits \cite{bertini2019exact, bertini2019entanglement, piroli2020exact, pavel2021correlations, claeys2021ergodic} and the Rule 54 quantum cellular automaton (QCA54)\cite{friedman2019integrable, klobas2021exact, klobas2021exact2, klobas2021ent}. The first two types are generically chaotic while QCA54 is integrable \cite{friedman2019integrable, gombor2021integrable, prosen2021reversible, gombor2022integrable}. In QCA54, relaxation dynamics and entanglement dynamics after a quantum quench from some initial states can be computed explicitly \cite{klobas2021exact, klobas2021exact2, klobas2021ent}.

Integrable circuits can be constructed by using $\check{R}$-matrices that satisfy the Yang-Baxter equation (YBE) as the local gates  (Fig.~\ref{fig1}).  Depending on the properties of the $\check{R}$-matrix, there are three possible cases: (i) classical stochastic cellular automata (SCA) if $\check{R}$ is stochastic; (ii) unitary QCA if $\check{R}$ is unitary; (iii) nonunitary QCA if
$\check{R}$ is completely positive and trace preserving (CPTP). Most of the previous work focused on the first and the second case, while integrable nonunitary QCA are less studied. 

Nonunitary quantum circuits \cite{farrelly2020review, brennen2003entanglement, piroli2020quantum}, especially unitary-projective ones \cite{li2018quantum, skinner2019meas, chan2019unitary, jian2020criticality, Ippoliti2021post, lu2021spacetime, Ippoliti2022fractal}, are important not only because they correspond to discrete-time versions of open quantum systems, but also since they can represent gate-based quantum computing and  can be used to study error-mitigation algorithms in near-term quantum devices.
One example of an integrable nonunitary QCA was recently obtained by trotterizing the Hubbard model with imaginary interaction strength \cite{Medvedyeva2016exact, sa2021integrable}. Such examples are highly nontrivial to build, which leaves a question whether there is a systematic way to construct integrable nonunitary quantum circuits. 

In this paper, we construct several more examples of integrable nonunitary quantum circuits by making use of the recent developments in the systematic construction of Yang-Baxter(YB) integrable Lindblad superoperators in the continuous-time case \cite{ziolkowska2020yang, de2021constructing}. The procedure is based on finding $\check{R}$-matrices that satisfy the YBE, such that their derivatives at special points correspond to integrable Lindblad superoperators. What we demonstrate here is that thus obtained $\check{R}$-matrices not only yield continuous modes, but can also be used as the building blocks of discrete-time nonunitary quantum circtuits.
In combination with the approach of Ref.  \cite{de2021constructing} this may pave the way for a more systematic construction of nonunitary integrable quantum circuits.

The rest of the paper is organized as follows: In Section \ref{sect2}, we review the construction of integrable nonunitary quantum circuits with two-site gates and continuous integrable Lindblad superoperators. By demonstrating the CPTP property of the $\check{R}$-matrices constructed from the latter, we show that these  $\check{R}$-matrices can be used to build integrable nonunitary quantum circuits. In Section \ref{sect3}, we apply the bond-site transformation to the quantum circuits with two-site gates to construct integrable nonunitary quantum circuits with three-site gates. In Section \ref{SCA}, we briefly discuss classical SCA when the density matrix is restricted to be diagonal. In particular, we show that the Rule 150 reversible cellular automaton (RCA150) \cite{bobenko1993two} can be embedded into one of the models discussed and the Rule 105 reversible cellular automaton (RCA105) can also be embedded if the regularity condition is relaxed. In the last section, we summarize and discuss a potential way to build integrable nonunitary quantum circuits systematically, and point the way for possible future directions. Some details are relegated to the appendixes.

\section{Integrable quantum circuits with two-site gates}
\label{sect2}
\subsection{Integrability structure}
\label{int_str}
In this section, we review the integrability structure of (two-site) translationally invariant quantum circuits with two-site gates \cite{vanicat2018integrable, sa2021integrable}. It is based on the standard quantum inverse-scattering method 
\cite{vsamaj2013introduction, eckle2019models, baxter2016exactly}. The central idea is to find a family of mutually commuting transfer matrices obtained from an $\check{R}$-matrix that satisfies the braid YBE. Then we can obtain a family of commuting conserved charges and even use algebraic Bethe ansatz techniques to obtain the spectrum of the transfer matrices and eigenfunctions.  For a quick introduction, see, e.g., Ref. \cite{Bertini2021finite}. 

\begin{figure}[t]
\centering
\includegraphics[width=0.4\textwidth]{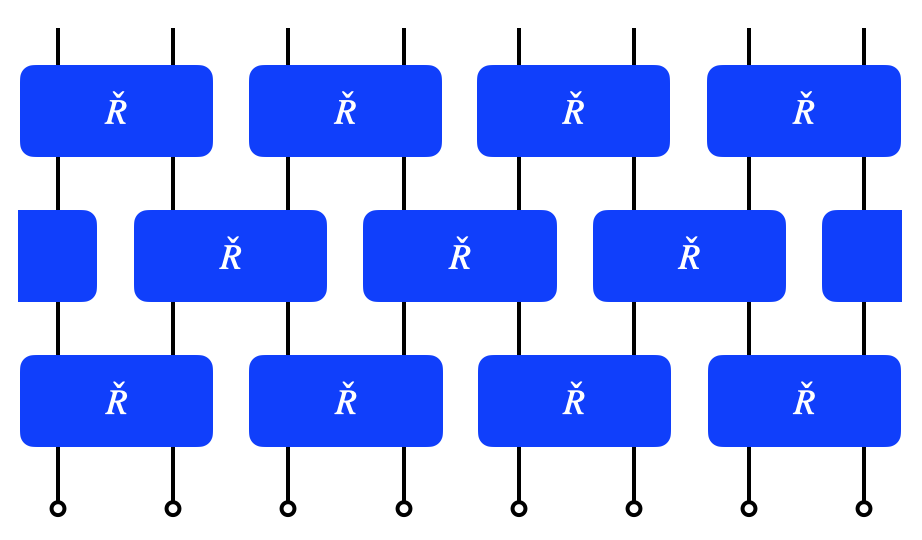}  
\caption{Floquet quantum circuit with period two. Each site has a Hilbert space of dimension $d$. Two-site gates $\check{R}$ that satisfy the YBE (Eq.(\ref{YBE})) act alternately.}
\label{fig1}
\end{figure}

Assume that we have an $\check{R}$-matrix, parametrized by $u$ and $v$, acting on two adjacent sites with local Hilbert space dimension $d$ each and that the $\check{R}$-matrix satisfies the braid YBE \cite{vsamaj2013introduction}:

\begin{align}
\check{R}_{23}(u, v)& \check{R}_{12}(u,w)\check{R}_{23}(v,w) =\nonumber \\
&\check{R}_{12}(v, w)\check{R}_{23}(u,w)\check{R}_{12}(u,v).
\label{YBE}
\end{align}
We let it act on each pair of sites $2j-1$ and $2j$ denoted by $(2j-1, 2j)$ with $j \in\{1, 2,..., L/2\}$ of a lattice of even length $L$ (Fig.~\ref{fig1}). We take periodic boundary conditions, namely, we identify $L+1\sim 1$. Then we let it act on all pairs $(2j, 2j+1)$ of sites for $j \in\{1, 2,..., L/2\}$. The combination of the actions is given by
\be
{\cal{V}}(u_1, u_2)  = \prod_{j=1}^{L/2}\check{R}_{2j, 2j+1} (u_1, u_2) \prod_{j=1}^{L/2}\check{R}_{2j-1, 2j} (u_1, u_2), 
\ee
which is called the Floquet operator.
Then we repeat the action. In this way, we construct a brickwork type Floquet  quantum circuit with period 2. If we take $d =2$ for a closed system, then $\check{R}$ is 4 by 4. If we are to study open systems, it is natural to use the density matrix formalism. Namely, the Floquet operator ${\cal{V}}$ acts on density matrix $\rho$. After vectorizing $\rho$, $\rho = \sum \rho_{ij} |i\rangle \langle j| \to |\rho\rangle=   \sum \rho_{ij} |i\rangle | j\rangle\rangle$, the local dimension becomes $d \to d^2$, and thus $\check{R}$ is 16 by 16.  In this section, we impose the regularity condition
\be \check{R}(u, u) = \1, \label{regcon}\ee 
or equivalently, \be R(u, u) \equiv P \check{R}(u, u) =P,\ee where $P$ is the permutation operator between two spaces defined by $P|i\rangle |j\rangle = |j\rangle |i\rangle$, which is also called the SWAP gate. It describes the trivial exchange process of the scattering of two particles \cite{vsamaj2013introduction}.  This condition, together with the YBE, implies \cite{gombor2021integrable} the inversion property, 
\be \check{R}_{12}(u, v) \check{R}_{12}(v, u) =\1.\label{inv}\ee 
 
To show that the circuit is YB integrable, we follow Ref.~\cite{sa2021integrable} and define a staggered transfer matrix:
\begin{align}
T(u) & =  \nonumber \\ 
&\Tr_{a}[R_{a L} (u, u_1 )R_{a L-1} (u, u_2 )... R_{a 2} (u, u_1 ) R_{a 1} (u, u_2 )]
\end{align}
where $a$ can be interpreted as an auxiliary space to take trace over. Then the YBE leads to 
$[T(u), T(v)]=0$ for any $u$ and $v$.

The regularity property and the inversion property then imply 
\begin{align}
 &T(u_1)   \nonumber \\
=&P_{12} P_{13}...P_{1L}   \check{R}_{12} (u_1, u_2 )\check{R}_{34} (u_1, u_2 )... \check{R}_{L-1 L} (u_1, u_2 ) \nonumber \\
 \equiv & U \check{R}_{12} (u_1, u_2 )\check{R}_{34} (u_1, u_2 )... \check{R}_{L-1 L} (u_1, u_2 ), 
\end{align} 
and
\begin{align}
& T(u_2) \nonumber \\  
=& P_{12} P_{13}...P_{1L} \check{R}_{23}^{-1} (u_1, u_2 )\check{R}_{45}^{-1} (u_1, u_2 )... \check{R}^{-1}_{L 1} (u_1, u_2 ) \nonumber \\
 \equiv & U  \check{R}_{23}^{-1} (u_1, u_2 )\check{R}_{45}^{-1} (u_1, u_2 )... \check{R}^{-1}_{L 1} (u_1, u_2 ),
\end{align}
where $U =P_{12} P_{13}...P_{1L}$ is a one-site shift. We have used the identity ${\cal{O}}_{ab} P_{ac} = P_{ac} {\cal{O}}_{cb}$ (for $b\neq c$) and $\Tr_a P_{a1} = \1$. Thus, 

\be
{\cal{V}} (u_1, u_2)  = T(u_2)^{-1} T(u_1)
\ee
and it commutes with $T(u)$ for any $u$. The commuting charges are given by
\be
Q_n^+ =\frac{d^n}{d u^n}  \ln T(u) |_{u =u_1}, \quad Q_n^- =\frac{d^n}{d u^n}  \ln T(u) |_{u =u_2}.   
\ee

\subsection{Integrable Lindblad Superoperators}\label{ILS}
Quantum circuits that we discussed may be regarded as coming from trotterization of  corresponding lattice models. If an open continuum system is in contact with a Markovian environment whose dynamics is much faster than  the internal one, the equation of motion is approximately given by the Lindbladian master equation 
\be
\dot{\rho} = {\cal{L}} \rho = i[\rho, H] + \sum_a l_a \rho l_a^{\dagger} - \frac{1}{2}\{ l_a^{\dagger} l_a, \rho\}, 
\label{lme}
\ee
where $\rho$ is the density operator, ${\cal{L}}$ is the Lindbladian and $l_a$ are jump operators. For a lattice system,  ${\cal{L}} = \sum_i {\cal{L}}_{i, i+1}$.

A recent study \cite{de2021constructing} constructed many new examples of integrable Lindblad superoperators following Ref.~\cite{de2019classifying, de2020classifying}.  We briefly summarize their approach. In terms of the vectorized density matrix, $\rho \in {\cal{H}}^{(1)} \otimes  {\cal{H}}^{(2)} $,  the Lindbladian density for a system with one family of jump operators has the explicit form
\begin{align}
{\cal{L}}_{i, i+1}(u) & = -i h_{i, i+1}^{(1)}(u) + i h_{i, i+1}^{(2)*}(u) +l_{i, i+1}^{(1)}(u) l_{i, i+1}^{(2)*}(u)  \nonumber \\  
& - \frac{1}{2} l_{i, i+1}^{(1)\dagger}(u) l_{i, i+1}^{(1)}(u) - \frac{1}{2} l_{i, i+1}^{(2)\text{T}}(u) l_{i, i+1}^{(2)*}(u).
\label{lin}
\end{align}
Here, ${\cal{H}}^{(1)}$ is the Hilbert space of two qubits and ${\cal{H}}^{(2)}$ is its canonical dual space; $h^{(s)}$ and $l^{(s)}$ act on the $s$-th Hilbert space. $^*$ denotes the complex conjugate and $^\text{T}$ the transpose. The dependence on the spectral parameter $u$ will be used in the following. 

To construct integrable ${\cal{L}}$, we use ansatze for ${\cal{L}}$ of the form in Eq.(\ref{lin}) and its associated $\check{R}$ satisfying
\be
\check{R}(u, u) =\1, \quad   {\cal{L}}_{i, i+1}(u) = \partial_v \check{R}_{i, i+1}(v, u)|_{v\to u}.
\label{initial}
\ee

Let $Q_n$ be the $n$-site conserved charge, in particular, $Q_2 = {\cal{L}}$. Higher charges are generated by 
\be
Q_{n+1}(u) = [ B(u), Q_n(u)], \quad n >1,
\label{q2}
\ee
where the boost operator $B(u):=  \partial_u + \sum_{i = -\infty}^{\infty} i {\cal{L}}_{i, i+1}(u)$. This expression is defined for infinite chains but reduces to finite chains consistently. Then impose the condition
\be [Q_2(u), Q_3(u)] =0\ee on the ansatz and solve the following equations obtained from the YBE:  
\begin{subequations}
\begin{gather}
[R_{13} R_{23}, {\cal{L}}_{12}(u)] = \dot{R}_{13} R_{23} - R_{13} \dot{R}_{23};\\
[R_{13} R_{12}, {\cal{L}}_{23}(v)] = R_{13} R'_{12} - R'_{13} R_{12}.
\end{gather}
\label{suther}
\end{subequations}  
In the first equation $R_{ij} = R_{ij}(u, u)$ and $\dot{R}$ is the derivative taken with respect to the first variable at $v=u$. In the second equation, $R_{ij} = R_{ij}(v, v)$ and the derivative is taken with respect to the second variable at $u =v$. Combining Eqs.(\ref{initial}), (\ref{q2}) and (\ref{suther}), we can obtain $\cal{L}$ and $\check{R}$. In the end, we can check if the $\check{R}$-matrix obtained this way satisfies the YBE (Eq.(\ref{YBE})).

Note that there are some degrees of freedom in $h$ and $l$ that keep ${\cal{L}}_{i, i+1}(u)$ invariant \cite{de2021constructing}, thus yielding the same $\check{R}$. However, different ${\cal{L}}_{i, i+1}(u)$ may give rise to $\check{R}$-matrices that are still equivalent via local basis transformations, reparameterizations, normalizations, discrete transformations, and twists. See Ref.\cite{de2019classifying, de2020classifying} for more information. In order to qualify as integrable non-unitary (open) quantum circuits,  the $\check{R}$-matrix has to be a quantum channel or, equivalently, satisfy CPTP \cite{preskill1998lecture}.

\subsection{CPTP}
To show that an $\check{R}$-matrix is CPTP, we can use the Choi-Jamiolkowski representation \cite{sa2021integrable} of $\check{R}$ and define the Choi matrix $D_{\beta \delta \zeta \theta}^{\alpha \gamma\varepsilon \eta} \equiv \check{R}_{\gamma \delta \eta \theta}^{\alpha \beta \varepsilon \zeta}$. Here, the first two (upper/lower) indices of $\check{R}$ are associated with the first site (after vectorization ($|i\rangle \langle j|\to |i\rangle |j\rangle $) while the last two associated with the second site. Each index can take two values. Consequently, $\check{R}$ is trace-preserving if and only if  $D^{\alpha \gamma \varepsilon\eta}_{\alpha \delta \varepsilon \theta} = \check{R}^{\alpha\alpha \varepsilon\varepsilon}_{\gamma \delta \eta \theta} = \delta_{\alpha}^{\gamma} \delta_{\theta}^{\eta}$; $\check{R}$ is completely positive if and only if $D$ is non-negative. Additionally,  $\check{R}$ is unital if and only if $D^{\alpha \gamma \varepsilon\eta}_{\beta \gamma \zeta \eta} = \check{R}^{\alpha\beta \varepsilon\zeta }_{\gamma \gamma \eta \eta} = \delta^{\alpha}_{\beta} \delta^{\varepsilon}_{\zeta}$. Unitality is not necessary for $\check{R}$ to be CPTP.  

It is easy to check explicitly that all $\check{R}$-matrices of all five cases in Ref.~\cite{de2021constructing} are CPTP. Additionally,  \be \tr(D (u))= \tr(D(0)) =4.
\ee
Hence, it has at most rank 2 and the two non-vanishing eigenvalues satisfy 
\be
\lambda_1 +\lambda_2 =4,
\ee 
and $\lambda_1, \lambda_2 \ge 0$ for $u\ge 0$. Since $\check{R}$ is CPTP, we have a Kraus representation \cite{preskill1998lecture} of  $\phi_{\beta \delta \zeta \theta}^{\alpha \gamma\varepsilon \eta} \equiv \check{R}_{\beta \zeta \delta \theta}^{\alpha \varepsilon \gamma \eta}$:
\be
\phi  =K_0\otimes K_0^* + K_1\otimes K_1^* , 
\ee
which satisfies 
\be
K_0^{\dagger}K_0 +K_1^{\dagger}K_1 = \1, 
\ee
where $^{\dagger}$ represents a Hermitian conjugate.
It acts on $\rho$ as
\be
\phi[\rho] = K_0\rho K_0^{\dagger} + K_1 \rho K_1^{\dagger}. 
\ee
$\check{R}$ is unital if 
\be
K_0 K_0^{\dagger} +K_1  K_1^{\dagger}= \1. 
\label{unital}
\ee

Thus, we can use these $\check{R}$-matrices to construct integrable nonunitary quantum circuits.
Note that the CPTP property of $\check{R}$ holds for all $u \ge 0$. However, when $u\to 0$, the quantum circuits reduce to the continuous systems described by the Lindbladian master equation (Eq.(\ref{lme})). Indeed, in a closed system with $d=2$, the $u\to 0$ limit of the quantum circuit described in Subsection \ref{int_str} reduces to $\lim_{N\to \infty} {\cal{V}}(u/N)^N  \to \exp(i u H)$, where $H$ can be interpreted as the hermitian Hamiltonian \cite{vanicat2018integrable}. Similarly, in the open system, when we take $u\to 0$, $ \lim_{N\to \infty} {\cal{V}}(u/N)^N \to \exp(u\La)$.
The quantum circuit reduces to a system with continuous `time' $u$ \cite{vanicat2018integrable}.

In general, if there are more than one family of jump operators, the expansion of the Kraus operators around $u =0$ is  \footnote{Note that if $\check{R}$ has a difference form, i.e., $\check{R}(u, v) = \check{R}(u-v)$, we simply write $\check{R}(u) = \check{R}(u, 0)$. Similarly for $K_a$ and other operators. In this case, $\La$ is independent on $u$.}
\begin{subequations}
    \begin{align}
K_0(u) & = I + (-i H - \frac{1}{2} \sum_{a >0} l_a^{\dagger} l_a ) u + O(u^2), \\
K_a (u) & = \sqrt{u} l_a  + O(u).
\end{align}
\label{kraus}
\end{subequations}

In the following we show that the five $\check{R}(u)$-matrices in  Ref.\cite{de2021constructing} constructed using integrable Lindblad superoperators are CPTP. Indeed, it would be interesting to see if all $\check{R}$-matrices constructed in this way are always CPTP; so far we have not been able to demonstrate that. 
In principle, we can also run the argument in reverse: If we have a YBE solution $\check{R}(u)$ that has the CPTP property, it can be used to obtain a corresponding integrable Lindblad superoperator.  

\subsection{Examples}
We list all five cases in 
Ref.~\cite{de2021constructing}. We can classify them into two categories. The first category includes Model B1 and Model B2 where two limiting cases ($u\to 0$ and $u\to \infty$) are both unitary. The rest (Model A1, Model A2, and Model B3) do not have this property.
\subsubsection{Model B1}
The (16 by 16) $R$-matrix we use is obtained from that given in Ref.~\cite{de2021constructing} by rescaling it by a factor $e^u/(u+1)$. $h =0$ and 
\be 
l = 
\bpm \tau & 0 & 0 & 0\\
0 & 0& 1 & 0 \\
0 &1 &0  & 0\\
0 & 0 & 0 & \kappa
\epm,
\ee
where $\tau = \pm 1$ and $\kappa =\pm 1$. For $\tau =\kappa =1$,  
\be
\check{R}(u) = \frac{\1 +u P}{u+1}, 
\ee
where $P$ is the permutation matrix. 
This is the $R$-matrix of the $SU(4)$ invariant chain if $u$ is taken to be purely imaginary. It is easy to see that $\check{R} = P R$ satisfies the inversion property Eq.(\ref{inv}).

In this case, the $\check{R}$-matrix is both CPTP and unital. The non-trivial eigenvalues of the Choi matrix $D$ are 
\be
\lambda_{\pm} = 2 \pm 2 \frac{\sqrt{u^2 -(\frac{3-\tau \kappa}{2} )u+1}}{u+1}. 
\ee
 
The Kraus operators are 
\be K_0(u) = \sqrt{\frac{1}{u+1}} \1_{4\times 4}, K_1(u) = \sqrt{\frac{u}{u+1}}
\bpm 1 & 0 & 0 & 0\\
0 & 0& \tau & 0 \\
0 &\tau &0  & 0\\
0 & 0 & 0 & \kappa\tau
\epm.
\ee
  
The action on $\rho$ is
\be
\phi[\rho] =  K_0 \rho K_0 +  K_1 \rho K_1 . 
\label{Kraus}
\ee
It is easy to check that  $K^{\dagger}_0 K_0 + K_1^{\dagger} K_1 =\1$.  When $u =0$, $K_0(0) = \1$ and $K_1(0) =0$. When $u\to \infty$, 
\be K_0(\infty) = 0, \quad K_1(\infty) = 
\bpm 1 & 0 & 0 & 0\\
0 & 0& \tau & 0 \\
0 &\tau &0  & 0\\
0 & 0 & 0 & \kappa\tau
\epm.
\label{k0k1}
\ee
Thus, the trivial unitary quantum circuit $\check{R} =\1$ is connected to the unitary circuit $U = K_1(\infty)$ by tuning $u$ from 0 to $\infty$.  For $\tau \kappa =1$, $K_1(\infty)$ corresponds to the XXZ model with $\Delta =\pm 1$. In this case, $K_1(\infty)$ will not generate entanglement for any (pure) product states, which is not true for the other two cases. For $\tau \kappa =- 1$, it corresponds to the XX model with a specific $h_z$. For all four cases, $K_1(\infty)$ is dual-unitary, and it belongs to the noninteracting case described in Ref. \cite{bertini2019exact}. 

That $\check{R}$ is unital implies that the maximally mixed state is a steady state. It is the quantum version of the bi-stochastic classical map \cite{pavel2021correlations}. The conserved charges can be obtained as in Ref.\cite{vanicat2018integrable}.

\subsubsection{Model B2}
This case includes the XX model with dephasing noise (which corresponds to the Hubbard model with imaginary coupling) discussed in Ref.~\cite{sa2021integrable} as the special case ($\phi =\pi/2$ in the following discussion). $h$ and $l$, respectively, are given below:
\begin{subequations}
\be 
h =2  \bpm 0 & 0 & 0 & 0\\
0 & 0& e^{\ii \phi} & 0 \\
0 &e^{-\ii \phi} &0  & 0\\
0 & 0 & 0 & 0
\epm,
\ee 
\begin{align}
&\frac{l}{\eta} = \nonumber \\
& \bpm \cosh(2u) & 0 & 0 & 0\\
0 & 1& \ii e^{\ii \phi} \sinh(2u) & 0 \\
0 & -\ii e^{-\ii \phi} \sinh(2u)& -1  & 0\\
0 & 0 & 0 & -\cosh(2u)
\epm,
\end{align}
\end{subequations}
where \be \eta (u) = \sqrt[4]{\frac{w^2}{ w^2\sinh^2 (2u) +1}}.\ee 

Define
\be 
\check{G}(u) =\frac{1}{a}  \bpm a & 0 & 0& 0\\ 0 & c &-\ii b e^{\ii \phi} & 0 \\  0 & -\ii b e^{-\ii \phi} & c& 0 \\ 0 & 0 & 0& a\epm,
\ee 
where \be a = \cosh (u), \quad b = \sinh(u), \quad c=1.\ee 
Also, take the matrix $e^{\beta}_{\alpha}$ whose matrix element is $(e^{\beta}_{\alpha})^{\beta'}_{\alpha'} =\delta_{\beta, \beta'}\delta_{\alpha, \alpha'}$, and
\begin{subequations}
\begin{gather} 
\check{r}_{\uparrow}(\lambda) = \check{G}^{\alpha \gamma}_{\beta \delta}(\lambda) e_{\alpha}^{\beta} \otimes \1_2 \otimes e_{\gamma}^{\delta} \otimes \1_2, \\
\check{r}_{\downarrow}(\lambda) = \check{G}^{\alpha \gamma}_{\beta \delta}(\lambda)\1_2\otimes e_{\alpha}^{\beta} \otimes \1_2 \otimes e_{\gamma}^{\delta}, 
\end{gather}
\end{subequations}
Define $\check{r} = \check{r}_{\uparrow} \check{r}^*_{\downarrow}$ and then the $\check{R}$-matrix can take the form
\be
\check{R}(u, v) = \beta \check{r}(u-v) + \alpha \check{r}(u+v) (\sigma^z \otimes \sigma^z \otimes \1_4),
\ee
where 
\be 
\beta = 
  \frac{\cosh(u - v) \cosh(
     \xi - \eta)}{(\cosh (u - v) \cosh( \xi - \eta) + \cosh(u + v) \sinh( \xi - \eta))},
\ee
$\alpha = 1-\beta$, and  
\be
\frac{\sinh(2\xi)}{\sinh(2u)} =\frac{\sinh(2\eta)}{\sinh(2v)} = w, \quad u\ge v.
\ee
The inversion condition, Eq.(\ref{inv}), is satisfied. $\check{R}(u, v)$ is not of difference form  and it is unital and CPTP. In particular, the nonvanishing eigenvalues of its Choi matrix are $4\alpha$ and $4\beta$. The Kraus operators are given by 
\be 
K_0 = \sqrt{\beta}   \check{G}(u-v),\quad  K_1 = \sqrt{\alpha}   \check{G}(u+v) (\sigma^z \otimes \1).
\ee 
When $u=v$, $\xi =\eta$, $\beta =1$ and $\alpha=0$. $K_0(0)=\1$ and $K_1(0) =0$. When $u \to \infty$, $\alpha =1$ and $\beta=0$. Then  
\be K_0(\infty) = 0, \quad K_1(\infty) = 
\bpm 1 & 0 & 0 & 0\\
0 & 0& -i e^{i \phi} & 0 \\
0 &i e^{-i \phi} &0  & 0\\
0 & 0 & 0 & -1
\epm.
\ee
Note that when $\phi =\pi/2$ this is same as Eq.(\ref{k0k1}) for $\tau=-1$ and $\kappa =-1$. Both limiting cases are unitary. The integrability of the circuit is also demonstrated numerically in Ref.\cite{sa2021integrable} by examining the spectral statistics. 

Note that both Model B1 and Model B2 have the following form
$\phi[\rho] = p U_0\rho U_0^{\dagger} + (1-p) U_1\rho U_1^{\dagger}$ with $0\le p \le 1$. It is compatible with the theorem that unital quantum channels can always be written as affine combinations of unitaries \cite{mendl2009unital}. In fact, a convex combination
of unitaries is a mixture of unitary channels.

\subsubsection{Model A1} 
In this model, $h$ and $l$ are given by 
\be 
h =\frac{1}{2}  \bpm 0 & 0 & 0 & 0\\
0 & 0& e^{\ii \phi} & 0 \\
0 &e^{-\ii \phi} &0  & 0\\
0 & 0 & 0 & 0
\epm,\quad
l=\\
 \bpm 0 & 0 & 0 & 0\\
0 & 0& 0& 0 \\
0 & 1& -i e^{-\ii \phi}  & 0\\
0 & 0 & 0 &0
\epm.
\ee   
The entries of the $R$-matrix are given in Ref. \cite{de2021constructing}. $\check{R}$ is CPTP  and satisfies the inversion condition, but is not unital in this case since Eq.(\ref{unital}) is not satisfied. The two nonvanishing eigenvalues of the Choi matrix are $
\lambda_{\pm} =2 \pm \sqrt{1 +2 e^{-2 u} + e^{-3 u}}$. Thus
$\lambda_{\pm} \ge 0$ when $u \ge 0$. 
The Kraus operators are given by 
\begin{subequations}
    \begin{align}
 K_0 = e^{-u/2}
\bpm 1 & 0 & 0 & 0\\
0 & e^{-u/2} & 0 & 0 \\
0 &0 &e^{u/2}  & 0\\
0 & 0 & 0 & 1 
\epm, \\
 K_1 =  \sqrt{1-e^{-u}} 
\bpm 1& 0 & 0 & 0\\
0 & e^{-u/2} & 0 & 0 \\
0 &-\ii e^{ \ii \phi} & 0 & 0\\
0 & 0 & 0 & 1
\epm.
\end{align}
\end{subequations}
The phases of $K_0$ and $K_1$ are irrelevant.  

\subsubsection{Model A2}
Here, $h$ and $l$ are given by 
\be 
h =\frac{\tau}{2}  \bpm 0 & 0 & 0 & 0\\
0 & 0& i  & 0 \\
0 &-i &0  & 0\\
0 & 0 & 0 & 0
\epm,\quad
l=\\
 \bpm 1 & 0 & 0 & 0\\
0 & 0& 0& 0 \\
0 & \tau & 1  & 0\\
1 & 0 & 0 &0
\epm,
\ee   
where $\tau =\pm 1$. 
The entries of the $R$-matrix of the Choi matrix are given in Ref. \cite{de2021constructing}. $\check{R}$ is CPTP  and satisfies the inversion condition, but is not unital in this case since Eq.(\ref{unital}) is not satisfied. 
The two nonvanishing eigenvalues are
$\lambda_{\pm} = 2 \pm 2 e^{-3u/2}$ and $\lambda_{\pm} \ge 0$ for $u \ge 0$. The Kraus operators are given by 
\begin{subequations}
    \begin{align}
    K_0= e^{-u} 
\bpm 1 & 0 & 0 & 0\\
0 & e^{u/2} & 0 & 0 \\
0 &-\tau (e^u-1)  &e^{u/2}  & 0\\
0 & 0 & 0 & e^{u} 
\epm, \\ K_1 = \sqrt{e^u -1} e^{-u} 
\bpm 1 & 0 & 0 & 0\\
0 & 0 & 0 & 0 \\
0 &\tau &e^{u/2}  & 0\\
e^{u/2} & 0 & 0 & 0 
\epm.
\end{align}
\end{subequations}
 $\check{R}$ is not unital in this case since Eq.(\ref{unital}) is not satisfied.

\subsubsection{Model B3}
$h$ and $l$ are given by 
\begin{subequations}
 \be
h  =\frac{1}{2}  \bpm 0 & 0 & 0 & 0\\
0 & 0& e^{\ii \phi} & 0 \\
0 &e^{-\ii \phi} &0  & 0\\
0 & 0 & 0 & 0
\epm,
\ee
\be
l =\sqrt{\frac{\gamma}{2}}
 \bpm \gamma & 0 & 0 & 0\\
0 & 1& i (\gamma-1) e^{i \phi}& 0 \\
0 & -i (\gamma+1) e^{-i \phi}& -1  & 0\\
0 & 0 & 0 &\gamma
\epm.
\ee
\end{subequations}
$\gamma\ge 0$. 
The entries of the $R$-matrix of the Choi matrix are given in Ref. \cite{de2021constructing}   \footnote{There are typos for the $R$-matrix of Model B3 in Ref. \cite{de2021constructing} : $R_4^{10} = e^{\ii \phi} 2\ii (\gamma+1)^2 \sinh(\alpha)/\zeta$ and $R_4^7 = -e^{-\ii \phi} 2\ii (\gamma+1)^2 \sinh(\alpha)/\zeta$.}. $\check{R}$ is CPTP  and satisfies the inversion condition, but is not unital for generic $\gamma$ in this case since Eq.(\ref{unital}) is not satisfied.  The Kraus operators are given by 
 
\begin{subequations}
    \begin{align}    
   &  K_0 = \nonumber \\
     & \frac{1}{\eta}
\bpm \eta  & 0 & 0 & 0\\
0 & 1 & -\ii (1-\gamma)\sinh (\alpha)e^{i \phi} & 0 \\
0 & -\ii (1+\gamma)\sinh (\alpha) e^{-i \phi} & 1 & 0\\
0 & 0 & 0 &  \eta 
\epm,
\end{align} 
\be  K_1 = \frac{\xi}{\eta}
\bpm 0 & 0 & 0 & 0\\
0 & -e^{-\alpha} (1-\gamma) & i(1-\gamma) e^{i \phi}& 0 \\
0 & i(1+\gamma)e^{-i \phi} &  e^{\alpha} (1+\gamma) & 0\\
0 & 0 & 0 & 0 
\epm,
\ee
\end{subequations}
where $\alpha  = (\gamma^2+1)u/2$, $\eta  = \gamma \sinh(\alpha) +\cosh(\alpha)$, and $\xi= \sqrt{\gamma \sinh(\alpha) /[(1+\gamma^2) \cosh(\alpha) +2\gamma \sinh(\alpha)] }$.

\section{Integrable quantum circuits with three-site gates}
\label{sect3}

In the previous section we showed how two-site $\check{R}$ matrices that appear in the integrable Lindbladian construction can be repurposed to build nonunitary quantum circuits. In this section, we extend this construction to integrable three-site nonunitary gates and circtuits by taking advantage of bond-site transformations. 

\subsection{Bond-site transformation}
\label{sec_bst}
A bond-site transformation (BST) maps a relative orientation of a pair of spin to a single  spin value.  We discuss BST following Ref.~\cite{gombor2021integrable}. For a 1D spin chain, we choose the computational basis $\mid s_1, s_2, ..., s_L\rangle$ where $s_i =\pm 1$. For each pair of neighboring spins, define a state $s'_i = s_i s_{i+1}$ on the bond. More concretely, if two neighboring sites are in the same state, then $s' =1$; if two neighbor sites are in different states,then  $s' =-1$. Thus, the state $s'=-1$ can be interpreted as the existence of a domain wall. Note that this is a two-to-one correspondence: states related by a global spin reflection are mapped to the same state on the bond. If open boundary conditions are imposed on the original spin chain and one of the boundary states is fixed, the states on the bonds after the BST are in one-to-one correspondence to the rest of the states. If periodic boundary conditions are taken, the mapping is two-to-one, and the bond states are constrained by $s'_1 s'_2\cdots s'_L =1$. Note that, similar to the Jordan-Wigner transformation \cite{fradkin2013field}, a BST is nonlocal because a domain wall is nonlocal in the original basis. Moreover, an entangled state can be mapped to an unentangled one. For example, the Greenberger–Horne–Zeilinger state $(\mid\uparrow\rangle^{\otimes L} +\mid\downarrow\rangle^{\otimes L})/\sqrt{2}$ is mapped to a product state.
 
On the operator level, an $(l+1)$-site operator may be mapped to an $l$-site operator after a BST. In particular, three-site operators that commute with the global spin reflection ($\Sigma^x_1\otimes \Sigma^x_2 \otimes \Sigma^x_3$) are in one-to-one correspondence with a two-site operators that commute with $\sigma^z \otimes \sigma^z$. 

\begin{figure}[H]
\centering
\includegraphics[width=0.2\textwidth]{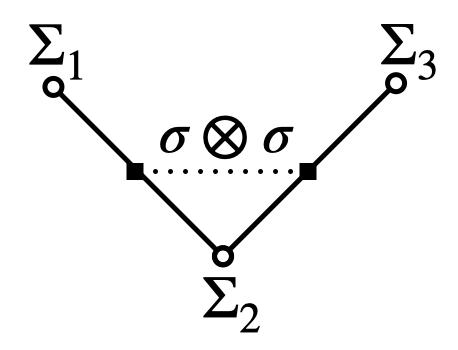}   
\label{figbts}
\end{figure}

They are listed below:  
\begin{align}
\sigma^x & \otimes \sigma^x  \leftrightarrow \1 \ \otimes \Sigma^x \otimes\ \1, \quad
\sigma^y \otimes \sigma^y  \leftrightarrow - \Sigma^z \otimes \Sigma^x \otimes \Sigma^z,\nonumber  \\
\sigma^z& \otimes \sigma^z  \leftrightarrow \Sigma^z \otimes\ \1 \ \otimes \Sigma^z, \quad
\sigma^y   \otimes \sigma^x  \leftrightarrow \Sigma^z \otimes \Sigma^y \otimes\ \1,\nonumber \\
\sigma^x &\otimes \sigma^y  \leftrightarrow\ \1 \  \otimes \Sigma^y \otimes \Sigma^z, \quad
\1   \otimes \sigma^z  \leftrightarrow -\1\ \otimes \Sigma^z \otimes \Sigma^z,\nonumber \\
\sigma^z &\otimes \1 \leftrightarrow -\Sigma^z \otimes \Sigma^z \otimes\ \1.
\end{align}
Consequently, general operators of the following form 
\begin{align}
\check{R} &= h_1 \1 \otimes \1 + \frac{h_2}{2} (\sigma^z \otimes \1 - \1 \otimes \sigma^z) + h_3 \sigma^+ \otimes \sigma^- \nonumber \\
&+ h_4 \sigma^- \otimes \sigma^+ + \frac{h_5}{2} (\sigma^z \otimes \1 + \1 \otimes \sigma^z) + h_6 \sigma^z \otimes \sigma^z \nonumber \\
&+ h_7 \sigma^-\otimes \sigma^- + h_8 \sigma^+ \otimes \sigma^+.
\end{align}
can all be mapped to three-site operators. 
Note that this is the general form of the ansatz used in Ref.~\cite{de2020classifying}. After a BST, the corresponding three-site operator $\check{G}$ satisfies
\be
[\check{G}_{123}, \check{G}_{345}]=0,
\label{comm}
\ee
since the operator acting on the common site (3) is either $\1$ or $\Sigma^z$. 

Since we want to consider open quantum systems, we need to generalize the transformation to mixed density matrices.
The vectorized density matrix, $|\rho\rangle$, is generated by $|s'_1,...\rangle |\tilde{s}'_1,...\rangle\rangle $. In particular, the basis states upon which the two-site gate $\check{R}$ discussed in the previous section acts are generated by $|s'_1, s'_2\rangle |\tilde{s}'_1, \tilde{s}'_2\rangle\rangle$.  Since we constructed $\check{R}$ using the Lindbladian of the form in Eq.(\ref{lin}), if both $h$ and $l$ commute with $\sigma^z\otimes \sigma^z$, the action of $\check{R}$ on $|s'_1, s'_2\rangle |\tilde{s}'_1, \tilde{s}'_2\rangle\rangle$ may preserve the product of $s'_1 s'_2$ and of $\tilde{s}'_1 \tilde{s}'_2$, separately. If this is the case, $|s'_1, s'_2\rangle$ and $ |\tilde{s}'_1, \tilde{s}'_2\rangle\rangle$ can separately be mapped to $|s_1, s_2, s_3\rangle$ and $|\tilde{s}_1, \tilde{s}_2, \tilde{s}_3\rangle\rangle$. There are four states corresponding to each $|s'_1, s'_2\rangle |\tilde{s}'_1, \tilde{s}'_2\rangle\rangle$ and the (64-dim) Hilbert space separates into four sectors. Moreover, $\check{R}$ can be regarded as a linear operator within each of the sectors. The corresponding three-site gate $\check{G}$ then also preserves each sector. Thus, $\check{G}$ is equal to the direct sum of four copies of $\check{R}$ when acting on the (64-dim) vectorized density operator, which, by linearity, implies that 
\begin{align}
\check{R}_{23}(u_1, u_2)&\check{R}_{12}(u_1, u_3) \check{R}_{23}(u_2, u_3) \nonumber \\ 
&=\check{R}_{12}(u_2, u_3)\check{R}_{23}(u_1, u_3)\check{R}_{12}(u_2, u_3). 
\label{RYBE}
\end{align}
leads to 
\begin{align}
\check{G}_{234}(u_1, u_2)& \check{G}_{123}(u_1, u_3) \check{G}_{234} (u_2, u_3) \nonumber \\
& = \check{G}_{123}(u_2, u_3) \check{G}_{234}(u_1, u_3) \check{G}_{123} (u_1, u_2).
\label{GYBE}
\end{align}
Also, if two-site (Kraus) operators $K_i$ satisfy
$ 
K_0^{\dagger}K_0 +K_1^{\dagger}K_1 = \1
$, the corresponding three-site operators $\tilde{K}_i = K_i\otimes K_i$ satisfy
$
\tilde{K}_0^{\dagger}\tilde{K}_0 +\tilde{K}_1^{\dagger}\tilde{K}_1 = \1. 
$ 

\begin{figure}[b]
\centering
\includegraphics[width=0.4\textwidth]{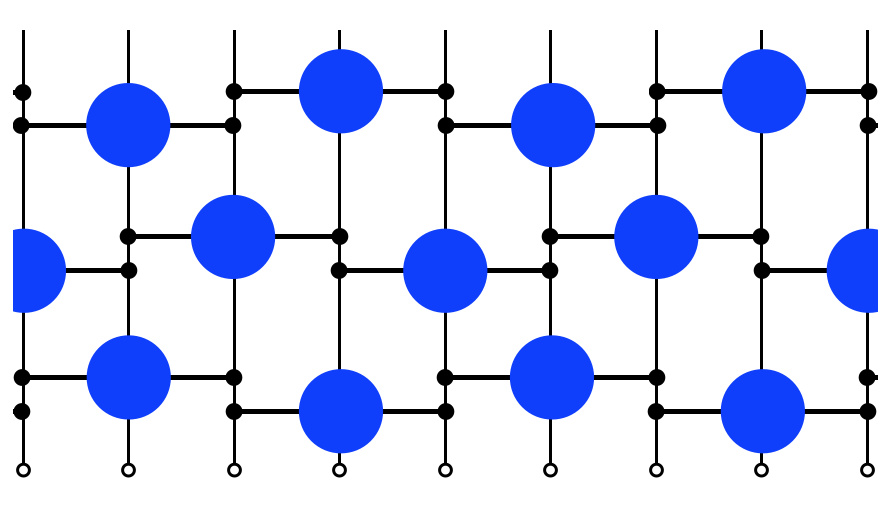}  
\caption{Floquet quantum circuit with staggered special three-site gates. A blue disk represents a general operator while a black dot can be $\1$ or $\Sigma^z$.}
\label{circuit3}
\end{figure}

\subsection{Integrability structure}
Since the BST is nonlocal, it is  not obvious whether all  local charges of a circuit with two-site gates are mapped to local charges of the corresponding circuit with three-site gates. To show that this is indeed the case, and thus the latter is integrable, we follow Ref. \cite{gombor2021integrable}.
 
We have already seen that if the two-site gate $\check{R}$ satisfies the YBE (Eq.(\ref{RYBE})) the corresponding three-site gate $\check{G}$ satisfies Eq.(\ref{comm}), Eq.(\ref{GYBE}), $ \check{G}_{123}(v, v)= \1$, and
\be \check{G}_{123}(u, v) \check{G}_{123}(v, u) = \1.\ee
Define $\check{L}_{123} (u) =\check{G}_{123}(u, 0)$ and 
\be 
\check{R}_{12, 34}(u,v) = \check{L}_{234}(u) \check{G}_{123}(u,v) \check{L}_{234}(v)^{-1}.
\label{rlgl}
\ee
Thus defined four-site $\check{R}$ then satisfies
\begin{align}
&\check{R}_{34, 56}(u_1, u_2) \check{R}_{12, 34}(u_1, u_3) \check{R}_{34, 56} (u_2, u_3) \nonumber \\
& = \check{R}_{12, 34}(u_2, u_3) \check{R}_{34, 56}(u_1, u_3) \check{R}_{12, 34} (u_1, u_2),
\end{align}
and we can construct a transfer matrix
\begin{align}
T(u) &= \Tr_{a,b} [R_{(1,2), (a,b)}(v, u)  R_{(3,4), (a,b)}(v, u)... \nonumber \\
& R_{(L-3,L-2), (a,b)}(v, u) R_{(L-1,L), (a,b)}(v, u)],
\end{align}
where $(a, b)$ denote the ``auxiliary space" and $R_{(a,b), (c,d)} \equiv P_{(a,b), (c,d)}\check{R}_{(a,b), (c,d)}$ ($P$ is the permutation operator).   
Regarding each pair of indices as a single index, we can follow the derivation Section \ref{int_str} to show that $ T(v)  = U^{-2}$, where $U$ is the one-site translation operator, and
\be [T(u), T(u')] =0.\ee 
Moreover, by Eq.(\ref{rlgl}),  $\check{R}_{12, 34}(u,0) = \check{L}_{234}(u) \check{L}_{123}(u)$. Then
\begin{align}
 T(0) &=  \nonumber \\
&\Tr_{a,b} [L_{1,2, b}L_{1,2, a}  L_{3,4, b} L_{3,4, a}\cdots   L_{L-1,L, b}L_{L-1,L, a}],
\end{align}
Define
\be {\cal{V}}(v, u) =  T(v)^{-1} T(u),\ee
then the Floquet operator (Fig.\ref{circuit3}) corresponds to 
\be {\cal{V}}(v, 0) = \prod_{j=1}^{L/2}\check{L}_{2j, 2j+1, 2j+2}(v)\prod_{j=1}^{L/2}\check{L}_{2j-1, 2j, 2j+1}(v).\ee 
The operator $ {\cal{V}}(v, u) $  leads to local conserved charges, the first being
\be 
Q_4  =\frac{d }{d u }  {\cal{V}}(v,u) |_{u =v}.
\ee 
It is a sum of four-site operators. Higher derivatives yield higher conserved charges. Therefore, we conclude that, if the BST is applicable to an integrable Floquet quantum circuit with two-site gates as discussed above, the corresponding circuit with three-site gates is also integrable. In particular, all five models discussed in the last section can be mapped to integrable circuits with nonunitary three-site gates.

\subsection{Example}
\label{example}
Let us discuss the simplest example:  Model B1 with $\tau \kappa =1$. Both Kraus operators of Model B1 are special cases of a general $\check{R}$-matrix of the zero-field eight-vertex (ZFEV) model:
\be
\check{R} =  \bpm a & 0 & 0& d\\ 0 & c & b& 0 \\  0 & b & c& 0 \\ d & 0 & 0& a\epm,
\label{R}
\ee 
where $a, b, c, d$ are usually functions of spectral parameters $u$ and $v$.  The $\check{R}$-matrix can also be written as  
\begin{align}
\check{R} & = \frac{1}{2} (a+c) \1 \otimes \1 +\frac{1}{2} (b+d)  \sigma^x  \otimes  \sigma^x \nonumber \\
&+\frac{1}{2} (b-d)  \sigma^y  \otimes  \sigma^y+ \frac{1}{2}( a- c) \sigma^z \otimes \sigma^z,
\label{R1}
\end{align}
which after a BST yields 
\begin{align}
\check{G} & = \frac{1}{2} (a+ d \Sigma_2^x)( 1 + \Sigma_1^z \Sigma_3^z)+  \frac{1}{2} (c+ b \Sigma_2^x)( 1- \Sigma_1^z \Sigma_3^z).
\label{G}
\end{align}
Similarly, the Kraus operators of Model B1 ($\tau \kappa =1$) are mapped to $K_0=  1/\sqrt{u+1} $ and 
\be
K_1=  \sqrt{\frac{u}{u+1}}[ \frac{1}{2} ( 1 + \Sigma_1^z \Sigma_3^z)+   \frac{1}{2}  \Sigma_2^x( 1- \Sigma_1^z \Sigma_3^z)].
\ee  
Their action on the transformed $\rho$ is similar to Eq.(\ref{Kraus}). In the limit $u \to \infty$, the action is unitary with $K_1$ the same as RCA150 (see Appendix \ref{app1}). Other cases in Ref. \cite{de2021constructing} can be described similarly.
 
\section{Classical SCA}
\label{SCA} 
In a standard cellular automaton \cite{Wolfram}, each site has one classical bit and it is updated according to a rule that takes the values of itself and its neighbors as the input \cite{farrelly2020review}. If the update rule is reversible, i.e., each output state has only one predecessor, it is also called a reversible cellular automaton; otherwise, it is irreversible.   Classical SCA can be obtained from QCA by constraining the density matrix $\rho$ to be diagonal, $\rho = \text{diag}(\rho_{\uparrow\uparrow}, \rho_{\uparrow\downarrow}, \rho_{\downarrow\uparrow}, \rho_{\downarrow\downarrow})$. That the restricted $\check{R}$ is stochastic (i.e., it maps a probability distribution to another) is guaranteed by the CPTP property of the original $\check{R}$. (A unital $\check{R}$ reduces to a bi-stochastic matrix \cite{pavel2021correlations}.) Preservation of the diagonal space leads to classical (discrete) flows of probabilities $P(s_1,s_2,..., s_L) = \langle s_1, s_2,...,s_L| \rho |s_1, s_2,...,s_L\rangle$ and the continuum limit of the five circuits with two-site classical gates is described Ref.\cite{de2021constructing}. As an example, Model B1 with $\tau \kappa =1$, for finite $u$ yields
\begin{align}
\text{diag}&(\rho_{\uparrow\uparrow}, \rho_{\uparrow\downarrow}, \rho_{\downarrow\uparrow}, \rho_{\downarrow\downarrow}) \nonumber \\
&\to \text{diag}(\rho_{\uparrow\uparrow}, \frac{\rho_{\uparrow\downarrow} +u \rho_{\downarrow\uparrow}}{u+1}, \frac{\rho_{\downarrow\uparrow}+u \rho_{\uparrow\downarrow}}{u+1}, \rho_{\downarrow\downarrow}).
\end{align}  
The corresponding transition matrix after a BST can be expressed as 
\begin{align}
\check{G}(u) & = \frac{1}{2}( 1 + \Sigma_1^z \Sigma_3^z)+  \frac{1}{2} \frac{1+ u \Sigma_2^x}{u+1}( 1- \Sigma_1^z \Sigma_3^z).
\end{align}

These integrable models are a subset of larger integrable family: ZFEV model (cf. Appendix \ref{app2}). Indeed, the application of a BST to the integrable chains Eq.(\ref{R1}) yields Eq.(\ref{G}). 
The coefficients are given in Eq.(\ref{abcd}). 
For $u \ge 0$ and $v \ge 0$, they are all nonnegative and $a + d = b + c =1$. They correspond to the ZFEV model with
\begin{align}
\Delta  &=  \frac{a^2 + b^2 - c^2 - d^2}{2(a b + c d)} = \text{sech} (2v), \nonumber \\
\Gamma & = \frac{a b - c d}{a b + c d} = \text{sech} (2v).
\end{align} 
Since $\text{sech}(2v) \le 1$, the ground state of the ZFEV model is always in the disordered phase or on its critical boundaries \cite{baxter2016exactly}. In particular, RCA150 corresponds to $v=0$ after a BST, i.e., it always sits on a critical line. For a recent analysis, see Ref.\cite{wilkinson2021exact}. For the ZFEV model, there are many symmetries which preserve the integrability \cite{baxter2016exactly}. In particular, $a\leftrightarrow d$ and $b\leftrightarrow c$ maps RCA150 to RCA105. Note that in Ref.\cite{gombor2021integrable}, RCA150 and RCA105 were embedded into two different families of integrable quantum models. This is because the regularity condition was imposed. We show in Appendix \ref{app3} that the regularity condition can be relaxed without breaking the integrability and it corresponds exactly to the $a\leftrightarrow d$ and $b\leftrightarrow c$ symmetry. The transition matrix is mapped to
\begin{align}
\check{G}(u) & = \frac{1}{2}( 1 - \Sigma_1^z \Sigma_3^z)+  \frac{1}{2} \frac{1+ u \Sigma_2^x}{u+1}( 1+ \Sigma_1^z \Sigma_3^z).
\end{align} 

Thus, we can see that some integrable SCA can be embedded into a family of integrable nonunitary QCA. In particular, RCA150 (and RCA105) can be embedded into families of integrable SCA, unitary QCA \cite{gombor2021integrable}, and nonunitary QCA.

\section{Conclusion and outlook}
We have constructed several two-site translationally invariant integrable nonunitary quantum circuits using the $\check{R}$-matrices that appeared in constructing integrable Lindblad superoperators in the continuous case  in Ref. \cite{de2021constructing}. The trotterized Hubbard model with imaginary interaction strength \cite{sa2021integrable} is one special case of our construction. We showed that these $\check{R}$-matrices generate integrable open systems by demonstrating that they are CPTP, and thus represent a valid quantum channel. By applying the BST, integrable quantum circuits with three-site gates can also be constructed.  Among them, the classical RCA150 (and RCA105) can be embedded into an integrable family describing quantum open systems. The classical discrete flows on the diagonal of density matrices of the entire family of Model B1 can be embedded into the well-known integrable ZFEV model with parameter restricted such that its ground state is always in the disordered phase (and its boundary critical lines). Since the BST is a special case of the more general Clifford transformations \cite{jones2021integrable}, other descendant integrable nonunitary quantum circuits can be similarly derived.  

In general, if we can find CPTP solutions to the YBE, we can use them directly to build integrable nonunitary quantum circuits. No Markovian approximation would be needed. Since we can regard the $\check{R}$-matrices of Model B1 and of Model B2 as those obtained from their unitary analogs via some analytical continuation, a closer look at the analytical continuation of known unitary solutions of YBE may help find other CPTP solutions. 

Our work points to a possible systematic way to obtain CPTP solutions to the YBE. Since all $\check{R}$-matrices of the five models in Ref.\cite{de2021constructing} are CPTP, a natural direction is to verify whether all $\check{R}$-matrices that can be constructed using the Lindbladian approach of Ref. \cite{de2021constructing} satisfy this property. In particular, if we use an ansatz of the form $\check{R}(u) =\1 + \sum_{n =1}^{\infty} r^{(n)}(\La) u^n$ where $r^{(n)}(\La)$ is a polynomial of degree $n$ in $\La$ \cite{de2019classifying}, the trace-preserving property of $\check{R}$ is automatically guaranteed since $\Tr[\La(\cdot)] =0$. Then it is sufficient to show that $\check{R}$ is completely positive.

A natural extension of the result in Ref.\cite{de2021constructing} to multiple families of jump operators may also lead to extension of the results in this paper to more general integrable open quantum circuits. In particular, it would be interesting to know if there is any connection to operator-space fragmentation discussed in Ref. \cite{Essler2020integrability} with two families of jump operators $l_{j, j+1}^{(1)} = \sqrt{J_1} \sigma_j^+\sigma_{j+1}^-$ and $l_{j, j+1}^{(2)} = \sqrt{J_2} \sigma_j^-\sigma_{j+1}^+$. 

We have noticed that  both Model B1 and Model B2 have the Kraus representation $\phi[\rho] =\sum_i \alpha_i U_i\rho U_i^{\dagger}$, a convex combination of unitary $U_i$.  It can be interpreted as an ensemble of quantum trajectories \cite{wiseman1996quantum, jian2020criticality}. If we focus on a single quantum trajectory and interpret nonunitary quantum gates as weak measurements following unitary evolution \cite{li2018quantum, Ippoliti2021post}, then we do not have to use vectorized density matrices. In particular, we can construct more integrable non-unitary quantum circuits by applying the spacetime duality to integrable unitary quantum circuits \cite{Ippoliti2021post}. All we need to do is relax the inversion property (Eq.(\ref{inv})) and regularity condition (Eq.(\ref{regcon})), which can be easily done. We will leave it for future work.

We have only proved the integrability of the open quantum circuits constructed. How the extensive number of conserved charges dictate the thermalization in integrable nonunitary quantum circuits is an important question. Explicit computation of time correlation functions can be challenging since the gates are no longer unitary. It would be interesting to see if the Bethe ansatz approach can yield useful information about the complex spectrum of the circuits \cite{ziolkowska2020yang} and, moreover, how dominant complex eigenvalues controls the long-time dynamics \cite{claeys2022correlations}. We may draw some inspirations from solutions to integrable systems with boundary drives \cite{vanicat2018integrable}. Numerically study of spectral statistics as in Ref.\cite{sa2021integrable} may provide another perspective. It is also helpful to compute dynamical entanglement-related qualities, such as entanglement spread, entanglement negativity, of the models. It will certainly give us a better understanding of irreversible/nonunitary QCA \cite{farrelly2020review, brennen2003entanglement, piroli2020quantum}.

\section*{Acknowledgement}
We thank Aashish Clerk for discussions. This material is based upon work supported by Laboratory Directed Research and Development (LDRD) funding from Argonne National Laboratory, provided by the Director, Office of Science, of the U.S. Department of Energy under Contract No. DE-AC02-06CH11357.

\appendix

\section{RCA150 and RCA105}
\label{app1}

RCA150 and RCA105 are two 3-to-1 cellular automata that preserve the spin reflection symmetry \cite{bobenko1993two,gombor2021integrable, wilkinson2021exact}. They belong to the category of ``exclusive one-spin facilitated" Fredrickson-Andersen (XOR-FA) models \cite{causer2020dynamics}.   Suppose that we are given three initial spins, $s_1, s_2, s_3 = \pm 1(\uparrow$/$\downarrow)$. We can determine $s_4$ (the final state of $s_1$) following the update rule.

\begin{figure}[H]
\centering
\includegraphics[width=0.1\textwidth]{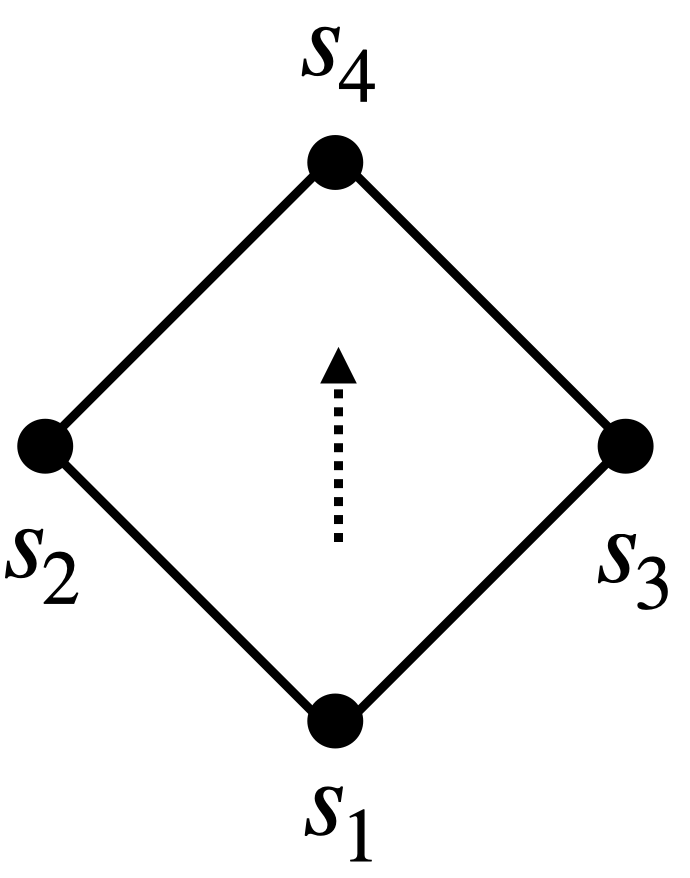}   
\label{150}
\end{figure} 
The update rule of RCA150 is given as follows: the central spin $s_1$ is flipped if its neighboring two spins, $s_2$ and $s_3$, are different or stay the same if they are equal. RCA105 is exactly the opposite: $s_4 =s_1$ if $s_2 \neq s_3$ and $s_4 = -s_1$ if $s_2 = s_3$.   RCA150 and RCA105 are not completely independent \cite{gombor2021integrable}. Indeed, let ${\cal{V}}_i^{(150)}$ and ${\cal{V}}_i^{(105)}$ be the product of update operators on the odd ($i =1$) or even ($i=2$) sublattice, repesctively. It is easy to see that
\be
{\cal{V}}_i^{(150)} = X_i {\cal{V}}_i^{(105)} = {\cal{V}}_i^{(105)} X_i,\quad i =1,2,   
\ee
where \be X_i = \prod_{j =1}^{L/2} \sigma^x_{2j+i}, \quad i= 1,2.\ee
Thus,
\begin{align}
 {\cal{V}}^{(105)} &= {\cal{V}}^{(105)}_1  {\cal{V}}^{(105)}_2 = X_1 {\cal{V}}^{(150)}_1  X_2{\cal{V}}^{(150)}_2 \nonumber \\
 &=X   {\cal{V}}^{(150)}_1  {\cal{V}}^{(150)}_2 = X {\cal{V}}^{(150)} . 
\label{eq_V}
\end{align} 
$X = X_1 X_2$. $X_i {\cal{V}}^{(150)}_i$ can serve as the new update rule.  Moreover, it is found in Ref.\cite{gombor2021integrable} that the conserved classical charge of RCA150 and RCA105 have the same form.
  
In Ref.\cite{gombor2021integrable}, RCA150 and RCA105 were embedded into two different models, respectively, and some sign differences were found. It turns out to be a result of imposing the regularity condition.  We will show in Appendix \ref{app3} that the regularity condition on $\check{R}$ is not necessary for a quantum circuit to be integrable and that, by relaxing the regularity condition, we may construct more integrable quantum circuits. In particular, RCA150 and RCA105  are related by a symmetry of the ZFEV  model.

\section{The ZFEV  model as a SCA}
\label{app2}
In this Appendix, we illustrate the relation between the classical SCA discussed in Section \ref{SCA} and the celebrated ZFEV model \cite{baxter2016exactly}. 

\begin{figure}[b]
\centering
\includegraphics[width=0.2\textwidth]{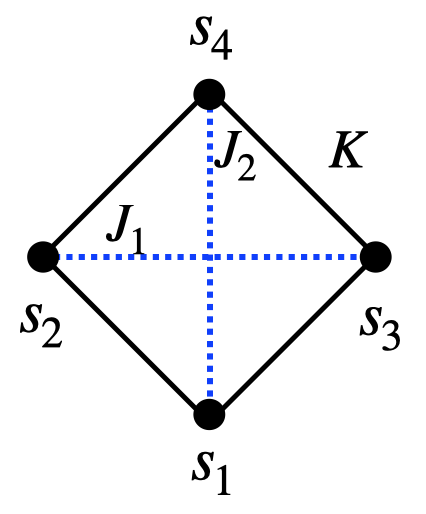}   

\caption{Couplings in a unit cell. $J_1$ and $J_2$ are the horizontal and vertical 2-spin interactions, respectively. $K$ is a 4-spin interaction.}
\label{is}
\end{figure} 

Let us first consider a 2D generalized Ising model on a square lattice with interactions in a unit cell given in Fig.(\ref{is}). It includes two spin-spin interactions, $J_1$ and $J_2$, and a 4-spin interaction, $K$. After absorbing $kT$ into the coupling constants, we can write the Boltzmann distribution of spins as a product of    
\be
E(s_4| s_3, s_2, s_1) = 
\exp [-( J_2 s_1 s_4 + J_1 s_2 s_3 + K s_1 s_2s_3s_4 +C)] 
\ee
where $C$ is a normalization constant and the product is taken over all unit cells. We used the notation $E(s_4| s_3, s_2, s_1)$ because the Ising model has an alternative interpretation as a 3-to-1 SCA as in Appendix \ref{app1}; namely, $s_1$ evolves into $s_4$ with a probability proportional to $E(s_4| s_3, s_2, s_1)$, provided $s_2$ and $s_3$ are given. To be more specific, use the basis $(s_2, s_3, s_1)$ and $(s_2, s_3, s_4)$ for the input and the output of the SCA and denote  

\be 
A =\bpm a & d\\   d & a  \epm, \quad
B =\bpm c & b\\ b & c \epm 
\label{eq_ising} 
\ee
where $ a = \exp(-J_1 -J_2 -K-C)$, $b =\exp(J_1 +J_2 -K-C)$,  $c =\exp(J_1 -J_2 +K-C)$, and $d= \exp(-J_1 +J_2 +K-C)$, and then the stochastic transitional matrix $P$ on a single square is given by  
\be P = \text{diag}(A, B, B, A). \ee
For $P$ to be a probability matrix, the following conditions must be satisfied: 
\be 
C = \frac{1}{2}[\ln (e^{J_2 +K} + e^{-J_2 -K}) +\ln (e^{-J_2 +K} + e^{J_2 -K})],
\ee
and
\be 
J_1 = \frac{1}{2}[\ln (e^{J_2 +K} + e^{-J_2 -K}) -\ln (e^{-J_2 +K} + e^{J_2 -K})].
\ee
Define 
\be
\tanh (u) = \exp(-2|J_2|),\quad \tanh (v) =\exp(-2|K|),
\ee
and then
\begin{align}
b &= \frac{\tanh(u)}{\tanh(u) + \tanh(v)}, \quad c = \frac{\tanh(v)}{\tanh(u) + \tanh(v)}, \nonumber \\
a &= \frac{1}{1+\tanh(u) \tanh(v)}, \quad d = \frac{\tanh(u) \tanh(v)} {1+\tanh(u) \tanh(v)},
\label{abcd}.
\end{align}

The map between the generalized Ising model and (two copies of) the ZFEV model is discussed in Ref. \cite{baxter2016exactly}. Indeed, define
\be 
\alpha_{ij} = s_{ij} s_{i, j+1},\quad \mu_{ij} =s_{ij} s_{i+1, j},
\ee
where $i, j$ represents the coordinate of a site on the lattice. (Note that 
this transformation is exactly the classical version of the BST discussed in Section \ref{sec_bst}.) Then the total energy (up to a constant) is 
\be 
{\cal{E}} = \sum_i \sum_j J_2 \alpha_{i+1, j} \mu_{ij} + J_1  \alpha_{ij} \mu_{ij} + K \alpha_{ij} \alpha_{i+1, j}
\ee 
with the constraint
\be
\mu_{ij} \alpha_{ij} \alpha_{i+1,j} \mu_{i, j+1}=1. 
\ee
Thus, each term corresponds to the Boltzmann weight of the vertex
\begin{figure}[H]
\centering
\includegraphics[width=0.1\textwidth]{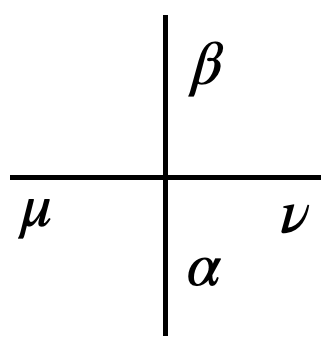}   
\end{figure} 
\noindent
which can be written as 
\be
w (\mu, \alpha| \beta, \nu) = \exp [- (J_2 \alpha \mu   +J_1 \beta \mu + K \alpha\beta)]
\ee
with $\mu \alpha \beta \nu =1$, and energies being
\begin{align}
\epsilon_1 &= \epsilon_2 = J_1 +J_2 +K,\quad \epsilon_3 = \epsilon_4 =- J_1 -J_2 +K \nonumber \\
\epsilon_5 &= \epsilon_6 = -J_1 +J_2 -K, \quad
\epsilon_7 = \epsilon_8 = J_1 -J_2 -K.
\end{align} 
Hence, the classical SCA with $a, b, c, d$ given above can be mapped to a ZFEV model (up to the boundary condition).

The $\check{R}$-matrix of the ZFEV model has the form 
\be 
\check{R} =  \bpm a & 0 & 0& d\\ 0 & c & b& 0 \\  0 & b & c& 0 \\ d & 0 & 0& a\epm,
\label{R3}
\ee 
and the family of commuting transfer matrices only depend on 
\be 
\Delta  =  \frac{a^2 + b^2 - c^2 - d^2}{2(a b + c d)} \ \text{and} \ 
\Gamma  = \frac{a b - c d}{a b + c d}.
\ee
One parametrization is given by 
\begin{align}
a &= \frac{\sn(u+v)}{\sn(u) +\sn(v)},  \quad b =  \frac{\sn(u)}{\sn(u) +\sn(v)}, \nonumber\\
c &= \frac{\sn(v)}{\sn(u) +\sn(v)},  \quad d =  \frac{k\sn(v)\sn(u) \sn(u+v)}{\sn(u) +\sn(v)},
\end{align}
where $\sn(u) = \sn(u; k)$ is a Jacobi elliptic function and $k\in [0,1]$. For $k=0$, $\text{sn}(u) = \sin(u)$; for $k=1$, $\text{sn}(u) = \tanh(u)$.
To associate the ZFEV model to a SCA, we impose the condition $a +d = b +c =1$. 
For $k=0$, we can take $u = \eta v$ and let $v\to 0$. Then
\be
a = 1, b = \frac{\eta}{\eta + 1}, c = \frac{1}{\eta + 1},  d = 0. 
\ee
For $k=1$, $a, b, c, d$ take the form in Eq.(\ref{abcd}).

Since $a +d = b +c =1$ and thus $|\Delta| = |\Gamma| \le 1$, the ZFEV model is always in the disordered phase (and/or on the critical lines)\cite{baxter2016exactly}. The ZFEV model have many symmetries  in terms of $a, b, c, d$. In particular, $a\leftrightarrow d$ and $b\leftrightarrow c$ does not breaking the integrability: $\check{R}' \equiv (\sigma^x\otimes \sigma^x) \check{R}$ still satisfies (the ``RLL" relation \cite{gombor2021integrable, Bertini2021finite}) $ 
\check{R}_{23}(u'')  \check{R}_{12}(u')\check{R}_{23}(u) = \check{R}_{12}(u)\check{R}_{23}(u')\check{R}_{12}(u'')
$ for some $u'' =u''(u, u')$. In the following appendix, we will show that it simply requires a relaxation of the regularity condition (Eq. (\ref{regcon})). There is also a duality in the ZFEV model that will map the system to an ordered phase \cite{baxter2016exactly}. However, the transition matrix is not always well-defined under such a transformation. 

\begin{figure}[b]
\centering
\includegraphics[width=0.48\textwidth]{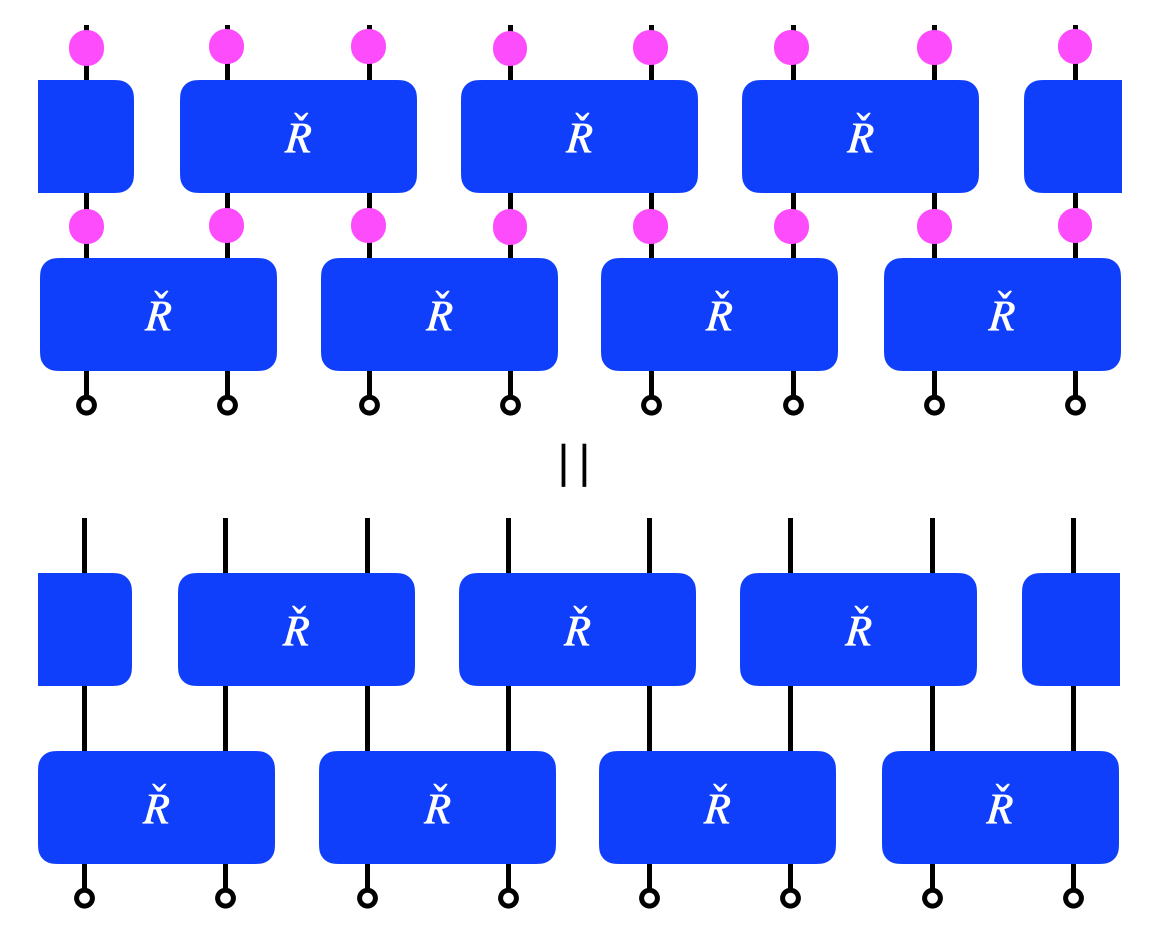}  
\caption{Floquet operator ${\cal{V}}'$ with insertion of  operators $U$ at each site after application of $\check{R}$ is equivalent to ${\cal{V}}$ if $[U\otimes U, \check{R}]=0$ and $U^2 =\1$..}
\label{reg2}
\end{figure}

\section{Relaxing the regularity condition}
\label{app3}

\begin{figure}[t]
\centering
\includegraphics[width=0.4\textwidth]{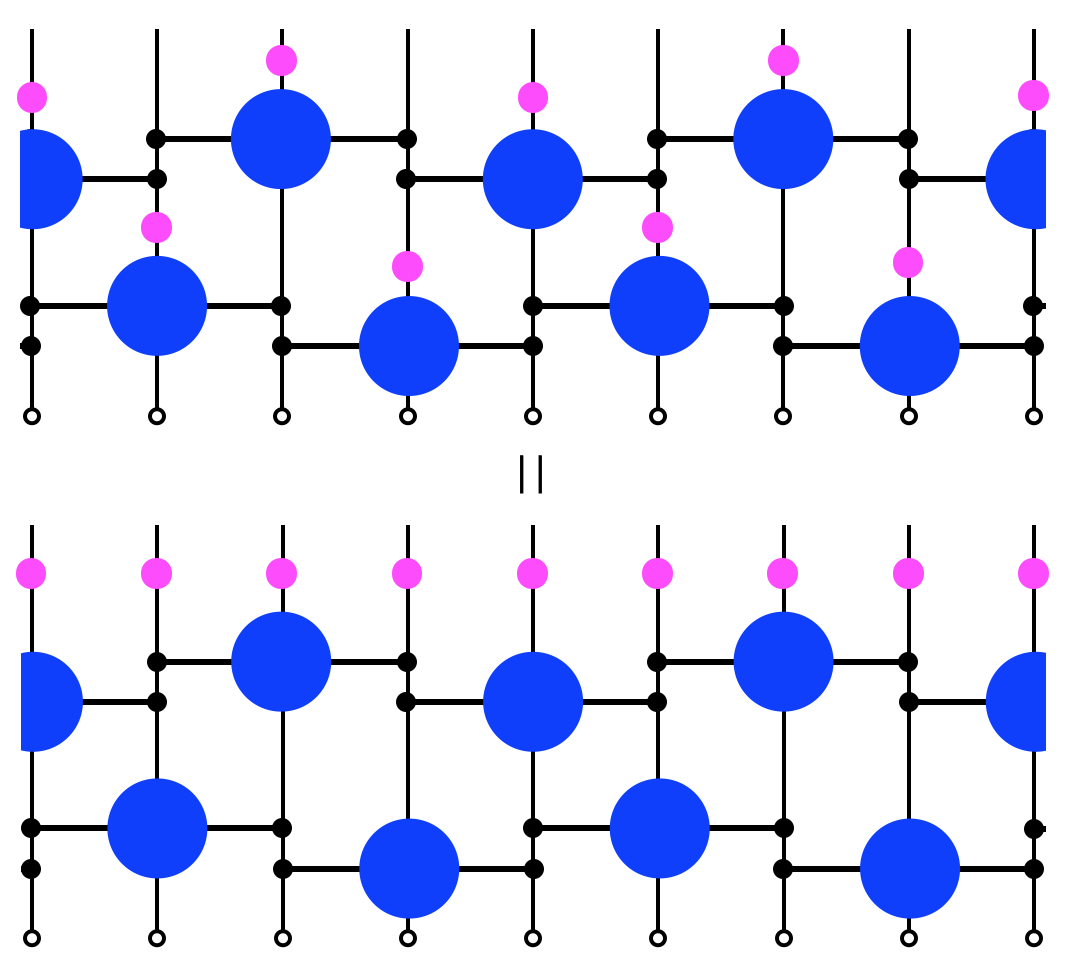}  
\caption{Floquet quantum circuit with update rule ${\cal{V}}'(v, 0)$. The magenta dot represents $\sigma^x$.}
\label{regcon3}
\end{figure}

We have shown in Section \ref{int_str} that the quantum circuit constructed from
\be
{\cal{V}}(\lambda_1, \lambda_2)  = \prod_{j=1}^{L/2}\check{R}_{2j, 2j+1} (\lambda_1, \lambda_2) \prod_{j=1}^{L/2}\check{R}_{2j-1, 2j} (\lambda_1, \lambda_2)
\label{vv}
\ee
is integrable if $\check{R}$ satisfies the YBE (Eq.(\ref{GYBE})) and the inversion property (Eq.(\ref{inv})). Let  $[U_i \otimes U_j, \check{R}_{ij}]=0$ and $U_i^2 =\1$.  Define $\check{R}'_{ij} \equiv (U_i \otimes U_j) \check{R}_{ij}$. Then 
\begin{align}
&{\cal{V}}'(\lambda_1, \lambda_2)  = \prod_{j=1}^{L/2}\check{R}'_{2j, 2j+1} (\lambda_1, \lambda_2) \prod_{j=1}^{L/2}\check{R}'_{2j-1, 2j} (\lambda_1, \lambda_2) \nonumber\\
 &= \prod_{j=1}^{L/2}\check{R}_{2j, 2j+1} (\lambda_1, \lambda_2) \prod_{j=1}^{L/2}\check{R}_{2j-1, 2j} (\lambda_1, \lambda_2) ={\cal{V}}(\lambda_1, \lambda_2). 
\end{align}
This identity is illustrated in Fig.~\ref{reg2}. Thus integrability is still guaranteed regardless of whether $\check{R}'$ satisfies the YBE. 

Take $d=2$, $U =\sigma^x$ and  $\check{R}$-matrix of the ZFEV model (Eq.(\ref{R3})). It is easy to see that $[\sigma^x \otimes \sigma^x , \check{R}]=0$, so Eq.~(\ref{vv}) holds after we multiply $\check{R}$ by $\sigma^x \otimes \sigma^x$ from left and integrability is preserved.  In fact,  it amounts to $a\leftrightarrow d$ and $b\leftrightarrow c$, and $\check{R}' = (\sigma^x\otimes \sigma^x) \check{R}$ still satisfies the YBE (more specifically, the ``RLL" relation \cite{gombor2021integrable, Bertini2021finite}). For a quantum spin chain with unitary $\check{R}$, this requires us to relax the regularity condition because now $\check{R}(u, u) =\sigma^x \otimes \sigma^x$. 

 After applying the BST, we have 
\begin{align}
&  {\cal{V}}'(v, 0) = \prod_{j=1}^{L/2}\check{L}'_{2j, 2j+1, 2j+2}(v)\prod_{j=1}^{L/2}\check{L}'_{2j-1, 2j, 2j+1}(v) \nonumber \\
& =X \prod_{j=1}^{L/2}\check{L}_{2j, 2j+1, 2j+2}(v)\prod_{j=1}^{L/2}\check{L}_{2j-1, 2j, 2j+1}(v) = X{\cal{V}}(v, 0). 
\end{align} 
where $X =\prod \sigma^x$. This relation is illustrated in Fig. \ref{regcon3}. 
The quantum circuit is integrable because $[X, T(u)]=0$ which implies that $ {\cal{V}}'(v, 0)$ commutes with $T(u)$ for any spectral parameter $u$. Note that this is exactly the relation between RCA150 and RCA105 as in Eq.(\ref{eq_V}). Thus, it is more natural to embed RCA105 into an integrable family by relaxing the regularity condition. It is consistent with the observation in Ref.\cite{gombor2021integrable} that the first nontrivial conserved classical charge of RCA150 and of RCA105 are the same. Moreover, there is no sign twist as in Ref.\cite{gombor2021integrable} for the corresponding quantum models.

\bibliographystyle{apsrev4-1}
\bibliography{QCA} 

\begin{thebibliography}{67}%
\makeatletter
\providecommand \@ifxundefined [1]{%
 \@ifx{#1\undefined}
}%
\providecommand \@ifnum [1]{%
 \ifnum #1\expandafter \@firstoftwo
 \else \expandafter \@secondoftwo
 \fi
}%
\providecommand \@ifx [1]{%
 \ifx #1\expandafter \@firstoftwo
 \else \expandafter \@secondoftwo
 \fi
}%
\providecommand \natexlab [1]{#1}%
\providecommand \enquote  [1]{``#1''}%
\providecommand \bibnamefont  [1]{#1}%
\providecommand \bibfnamefont [1]{#1}%
\providecommand \citenamefont [1]{#1}%
\providecommand \href@noop [0]{\@secondoftwo}%
\providecommand \href [0]{\begingroup \@sanitize@url \@href}%
\providecommand \@href[1]{\@@startlink{#1}\@@href}%
\providecommand \@@href[1]{\endgroup#1\@@endlink}%
\providecommand \@sanitize@url [0]{\catcode `\\12\catcode `\$12\catcode
  `\&12\catcode `\#12\catcode `\^12\catcode `\_12\catcode `\%12\relax}%
\providecommand \@@startlink[1]{}%
\providecommand \@@endlink[0]{}%
\providecommand \url  [0]{\begingroup\@sanitize@url \@url }%
\providecommand \@url [1]{\endgroup\@href {#1}{\urlprefix }}%
\providecommand \urlprefix  [0]{URL }%
\providecommand \Eprint [0]{\href }%
\providecommand \doibase [0]{http://dx.doi.org/}%
\providecommand \selectlanguage [0]{\@gobble}%
\providecommand \bibinfo  [0]{\@secondoftwo}%
\providecommand \bibfield  [0]{\@secondoftwo}%
\providecommand \translation [1]{[#1]}%
\providecommand \BibitemOpen [0]{}%
\providecommand \bibitemStop [0]{}%
\providecommand \bibitemNoStop [0]{.\EOS\space}%
\providecommand \EOS [0]{\spacefactor3000\relax}%
\providecommand \BibitemShut  [1]{\csname bibitem#1\endcsname}%
\let\auto@bib@innerbib\@empty
\bibitem [{\citenamefont {Deutsch}(1991)}]{deutsch1991quantum}%
  \BibitemOpen
  \bibfield  {author} {\bibinfo {author} {\bibfnamefont {J.~M.}\ \bibnamefont
  {Deutsch}},\ }\href {\doibase 10.1103/PhysRevA.43.2046} {\bibfield  {journal}
  {\bibinfo  {journal} {Phys. Rev. A}\ }\textbf {\bibinfo {volume} {43}},\
  \bibinfo {pages} {2046} (\bibinfo {year} {1991})}\BibitemShut {NoStop}%
\bibitem [{\citenamefont {Srednicki}(1994)}]{srednicki1994chaos}%
  \BibitemOpen
  \bibfield  {author} {\bibinfo {author} {\bibfnamefont {M.}~\bibnamefont
  {Srednicki}},\ }\href {\doibase 10.1103/PhysRevE.50.888} {\bibfield
  {journal} {\bibinfo  {journal} {Phys. Rev. E}\ }\textbf {\bibinfo {volume}
  {50}},\ \bibinfo {pages} {888} (\bibinfo {year} {1994})}\BibitemShut
  {NoStop}%
\bibitem [{\citenamefont {Moudgalya}\ \emph {et~al.}(2021)\citenamefont
  {Moudgalya}, \citenamefont {Bernevig},\ and\ \citenamefont
  {Regnault}}]{moudgalya2021quantum}%
  \BibitemOpen
  \bibfield  {author} {\bibinfo {author} {\bibfnamefont {S.}~\bibnamefont
  {Moudgalya}}, \bibinfo {author} {\bibfnamefont {B.~A.}\ \bibnamefont
  {Bernevig}}, \ and\ \bibinfo {author} {\bibfnamefont {N.}~\bibnamefont
  {Regnault}},\ }\href {https://arxiv.org/abs/2109.00548} {\bibfield  {journal}
  {\bibinfo  {journal} {arXiv:2109.00548}\ } (\bibinfo {year}
  {2021})}\BibitemShut {NoStop}%
\bibitem [{\citenamefont {Polkovnikov}\ \emph {et~al.}(2011)\citenamefont
  {Polkovnikov}, \citenamefont {Sengupta}, \citenamefont {Silva},\ and\
  \citenamefont {Vengalattore}}]{Polkovnikov2011}%
  \BibitemOpen
  \bibfield  {author} {\bibinfo {author} {\bibfnamefont {A.}~\bibnamefont
  {Polkovnikov}}, \bibinfo {author} {\bibfnamefont {K.}~\bibnamefont
  {Sengupta}}, \bibinfo {author} {\bibfnamefont {A.}~\bibnamefont {Silva}}, \
  and\ \bibinfo {author} {\bibfnamefont {M.}~\bibnamefont {Vengalattore}},\
  }\href {\doibase 10.1103/RevModPhys.83.863} {\bibfield  {journal} {\bibinfo
  {journal} {Rev. Mod. Phys.}\ }\textbf {\bibinfo {volume} {83}},\ \bibinfo
  {pages} {863} (\bibinfo {year} {2011})}\BibitemShut {NoStop}%
\bibitem [{\citenamefont {Nandkishore}\ and\ \citenamefont
  {Huse}(2015)}]{nandkishore2015many}%
  \BibitemOpen
  \bibfield  {author} {\bibinfo {author} {\bibfnamefont {R.}~\bibnamefont
  {Nandkishore}}\ and\ \bibinfo {author} {\bibfnamefont {D.~A.}\ \bibnamefont
  {Huse}},\ }\href {\doibase 10.1146/annurev-conmatphys-031214-014726}
  {\bibfield  {journal} {\bibinfo  {journal} {Annu. Rev. Condens. Matter
  Phys.}\ }\textbf {\bibinfo {volume} {6}},\ \bibinfo {pages} {15} (\bibinfo
  {year} {2015})}\BibitemShut {NoStop}%
\bibitem [{\citenamefont {D'Alessio}\ \emph {et~al.}(2016)\citenamefont
  {D'Alessio}, \citenamefont {Kafri}, \citenamefont {Polkovnikov},\ and\
  \citenamefont {Rigol}}]{d2016quantum}%
  \BibitemOpen
  \bibfield  {author} {\bibinfo {author} {\bibfnamefont {L.}~\bibnamefont
  {D'Alessio}}, \bibinfo {author} {\bibfnamefont {Y.}~\bibnamefont {Kafri}},
  \bibinfo {author} {\bibfnamefont {A.}~\bibnamefont {Polkovnikov}}, \ and\
  \bibinfo {author} {\bibfnamefont {M.}~\bibnamefont {Rigol}},\ }\href
  {\doibase 10.1080/00018732.2016.1198134} {\bibfield  {journal} {\bibinfo
  {journal} {Adv. Phys.}\ }\textbf {\bibinfo {volume} {65}},\ \bibinfo {pages}
  {239} (\bibinfo {year} {2016})}\BibitemShut {NoStop}%
\bibitem [{\citenamefont {Abanin}\ \emph {et~al.}(2019)\citenamefont {Abanin},
  \citenamefont {Altman}, \citenamefont {Bloch},\ and\ \citenamefont
  {Serbyn}}]{Abanin2019colloquium}%
  \BibitemOpen
  \bibfield  {author} {\bibinfo {author} {\bibfnamefont {D.~A.}\ \bibnamefont
  {Abanin}}, \bibinfo {author} {\bibfnamefont {E.}~\bibnamefont {Altman}},
  \bibinfo {author} {\bibfnamefont {I.}~\bibnamefont {Bloch}}, \ and\ \bibinfo
  {author} {\bibfnamefont {M.}~\bibnamefont {Serbyn}},\ }\href {\doibase
  10.1103/RevModPhys.91.021001} {\bibfield  {journal} {\bibinfo  {journal}
  {Rev. Mod. Phys.}\ }\textbf {\bibinfo {volume} {91}},\ \bibinfo {pages}
  {021001} (\bibinfo {year} {2019})}\BibitemShut {NoStop}%
\bibitem [{\citenamefont {Castro-Alvaredo}\ \emph {et~al.}(2016)\citenamefont
  {Castro-Alvaredo}, \citenamefont {Doyon},\ and\ \citenamefont
  {Yoshimura}}]{Castro2016emergent}%
  \BibitemOpen
  \bibfield  {author} {\bibinfo {author} {\bibfnamefont {O.~A.}\ \bibnamefont
  {Castro-Alvaredo}}, \bibinfo {author} {\bibfnamefont {B.}~\bibnamefont
  {Doyon}}, \ and\ \bibinfo {author} {\bibfnamefont {T.}~\bibnamefont
  {Yoshimura}},\ }\href {\doibase 10.1103/PhysRevX.6.041065} {\bibfield
  {journal} {\bibinfo  {journal} {Phys. Rev. X}\ }\textbf {\bibinfo {volume}
  {6}},\ \bibinfo {pages} {041065} (\bibinfo {year} {2016})}\BibitemShut
  {NoStop}%
\bibitem [{\citenamefont {Bertini}\ \emph {et~al.}(2016)\citenamefont
  {Bertini}, \citenamefont {Collura}, \citenamefont {De~Nardis},\ and\
  \citenamefont {Fagotti}}]{Bertini2016transport}%
  \BibitemOpen
  \bibfield  {author} {\bibinfo {author} {\bibfnamefont {B.}~\bibnamefont
  {Bertini}}, \bibinfo {author} {\bibfnamefont {M.}~\bibnamefont {Collura}},
  \bibinfo {author} {\bibfnamefont {J.}~\bibnamefont {De~Nardis}}, \ and\
  \bibinfo {author} {\bibfnamefont {M.}~\bibnamefont {Fagotti}},\ }\href
  {\doibase 10.1103/PhysRevLett.117.207201} {\bibfield  {journal} {\bibinfo
  {journal} {Phys. Rev. Lett.}\ }\textbf {\bibinfo {volume} {117}},\ \bibinfo
  {pages} {207201} (\bibinfo {year} {2016})}\BibitemShut {NoStop}%
\bibitem [{\citenamefont {Bertini}\ \emph {et~al.}(2021)\citenamefont
  {Bertini}, \citenamefont {Heidrich-Meisner}, \citenamefont {Karrasch},
  \citenamefont {Prosen}, \citenamefont {Steinigeweg},\ and\ \citenamefont
  {\ifmmode \check{Z}\else \v{Z}\fi{}nidari\ifmmode~\check{c}\else
  \v{c}\fi{}}}]{Bertini2021finite}%
  \BibitemOpen
  \bibfield  {author} {\bibinfo {author} {\bibfnamefont {B.}~\bibnamefont
  {Bertini}}, \bibinfo {author} {\bibfnamefont {F.}~\bibnamefont
  {Heidrich-Meisner}}, \bibinfo {author} {\bibfnamefont {C.}~\bibnamefont
  {Karrasch}}, \bibinfo {author} {\bibfnamefont {T.}~\bibnamefont {Prosen}},
  \bibinfo {author} {\bibfnamefont {R.}~\bibnamefont {Steinigeweg}}, \ and\
  \bibinfo {author} {\bibfnamefont {M.}~\bibnamefont {\ifmmode \check{Z}\else
  \v{Z}\fi{}nidari\ifmmode~\check{c}\else \v{c}\fi{}}},\ }\href {\doibase
  10.1103/RevModPhys.93.025003} {\bibfield  {journal} {\bibinfo  {journal}
  {Rev. Mod. Phys.}\ }\textbf {\bibinfo {volume} {93}},\ \bibinfo {pages}
  {025003} (\bibinfo {year} {2021})}\BibitemShut {NoStop}%
\bibitem [{\citenamefont {Alba}\ \emph {et~al.}(2021)\citenamefont {Alba},
  \citenamefont {Bertini}, \citenamefont {Fagotti}, \citenamefont {Piroli},\
  and\ \citenamefont {Ruggiero}}]{Alba2021genearlized}%
  \BibitemOpen
  \bibfield  {author} {\bibinfo {author} {\bibfnamefont {V.}~\bibnamefont
  {Alba}}, \bibinfo {author} {\bibfnamefont {B.}~\bibnamefont {Bertini}},
  \bibinfo {author} {\bibfnamefont {M.}~\bibnamefont {Fagotti}}, \bibinfo
  {author} {\bibfnamefont {L.}~\bibnamefont {Piroli}}, \ and\ \bibinfo {author}
  {\bibfnamefont {P.}~\bibnamefont {Ruggiero}},\ }\href {\doibase
  10.1088/1742-5468/ac257d} {\bibfield  {journal} {\bibinfo  {journal} {J.
  Stat. Mech.}\ }\textbf {\bibinfo {volume} {2021}},\ \bibinfo {pages} {114004}
  (\bibinfo {year} {2021})}\BibitemShut {NoStop}%
\bibitem [{\citenamefont {Schollw\"ock}(2005)}]{schollwock2005density}%
  \BibitemOpen
  \bibfield  {author} {\bibinfo {author} {\bibfnamefont {U.}~\bibnamefont
  {Schollw\"ock}},\ }\href {\doibase 10.1103/RevModPhys.77.259} {\bibfield
  {journal} {\bibinfo  {journal} {Rev. Mod. Phys.}\ }\textbf {\bibinfo {volume}
  {77}},\ \bibinfo {pages} {259} (\bibinfo {year} {2005})}\BibitemShut
  {NoStop}%
\bibitem [{\citenamefont {Schollw{\"o}ck}(2011)}]{schollwock2011density}%
  \BibitemOpen
  \bibfield  {author} {\bibinfo {author} {\bibfnamefont {U.}~\bibnamefont
  {Schollw{\"o}ck}},\ }\href {\doibase 10.1016/j.aop.2010.09.012} {\bibfield
  {journal} {\bibinfo  {journal} {Ann. Phys.}\ }\textbf {\bibinfo {volume}
  {326}},\ \bibinfo {pages} {96} (\bibinfo {year} {2011})}\BibitemShut
  {NoStop}%
\bibitem [{\citenamefont {Pekker}\ \emph {et~al.}(2014)\citenamefont {Pekker},
  \citenamefont {Refael}, \citenamefont {Altman}, \citenamefont {Demler},\ and\
  \citenamefont {Oganesyan}}]{Pekker2014hilbert}%
  \BibitemOpen
  \bibfield  {author} {\bibinfo {author} {\bibfnamefont {D.}~\bibnamefont
  {Pekker}}, \bibinfo {author} {\bibfnamefont {G.}~\bibnamefont {Refael}},
  \bibinfo {author} {\bibfnamefont {E.}~\bibnamefont {Altman}}, \bibinfo
  {author} {\bibfnamefont {E.}~\bibnamefont {Demler}}, \ and\ \bibinfo {author}
  {\bibfnamefont {V.}~\bibnamefont {Oganesyan}},\ }\href {\doibase
  10.1103/PhysRevX.4.011052} {\bibfield  {journal} {\bibinfo  {journal} {Phys.
  Rev. X}\ }\textbf {\bibinfo {volume} {4}},\ \bibinfo {pages} {011052}
  (\bibinfo {year} {2014})}\BibitemShut {NoStop}%
\bibitem [{\citenamefont {Cirac}\ \emph {et~al.}(2021)\citenamefont {Cirac},
  \citenamefont {P\'erez-Garc\'{\i}a}, \citenamefont {Schuch},\ and\
  \citenamefont {Verstraete}}]{cirac2021matrix}%
  \BibitemOpen
  \bibfield  {author} {\bibinfo {author} {\bibfnamefont {J.~I.}\ \bibnamefont
  {Cirac}}, \bibinfo {author} {\bibfnamefont {D.}~\bibnamefont
  {P\'erez-Garc\'{\i}a}}, \bibinfo {author} {\bibfnamefont {N.}~\bibnamefont
  {Schuch}}, \ and\ \bibinfo {author} {\bibfnamefont {F.}~\bibnamefont
  {Verstraete}},\ }\href {\doibase 10.1103/RevModPhys.93.045003} {\bibfield
  {journal} {\bibinfo  {journal} {Rev. Mod. Phys.}\ }\textbf {\bibinfo {volume}
  {93}},\ \bibinfo {pages} {045003} (\bibinfo {year} {2021})}\BibitemShut
  {NoStop}%
\bibitem [{\citenamefont {Baxter}(2016)}]{baxter2016exactly}%
  \BibitemOpen
  \bibfield  {author} {\bibinfo {author} {\bibfnamefont {R.~J.}\ \bibnamefont
  {Baxter}},\ }\href@noop {} {\emph {\bibinfo {title} {Exactly solved models in
  statistical mechanics}}}\ (\bibinfo  {publisher} {Elsevier},\ \bibinfo {year}
  {2016})\BibitemShut {NoStop}%
\bibitem [{\citenamefont {{\v{S}}amaj}\ and\ \citenamefont
  {Bajnok}(2013)}]{vsamaj2013introduction}%
  \BibitemOpen
  \bibfield  {author} {\bibinfo {author} {\bibfnamefont {L.}~\bibnamefont
  {{\v{S}}amaj}}\ and\ \bibinfo {author} {\bibfnamefont {Z.}~\bibnamefont
  {Bajnok}},\ }\href@noop {} {\emph {\bibinfo {title} {Introduction to the
  statistical physics of integrable many-body systems}}}\ (\bibinfo
  {publisher} {Cambridge University Press},\ \bibinfo {year}
  {2013})\BibitemShut {NoStop}%
\bibitem [{\citenamefont {Eckle}(2019)}]{eckle2019models}%
  \BibitemOpen
  \bibfield  {author} {\bibinfo {author} {\bibfnamefont {H.-P.}\ \bibnamefont
  {Eckle}},\ }\href@noop {} {\emph {\bibinfo {title} {Models of Quantum Matter:
  A First Course on Integrability and the Bethe Ansatz}}}\ (\bibinfo
  {publisher} {Oxford University Press},\ \bibinfo {year} {2019})\BibitemShut
  {NoStop}%
\bibitem [{\citenamefont {Granet}\ and\ \citenamefont
  {Essler}(2021)}]{Granet2021systematic}%
  \BibitemOpen
  \bibfield  {author} {\bibinfo {author} {\bibfnamefont {E.}~\bibnamefont
  {Granet}}\ and\ \bibinfo {author} {\bibfnamefont {F.~H.~L.}\ \bibnamefont
  {Essler}},\ }\href {\doibase 10.21468/SciPostPhys.11.3.068} {\bibfield
  {journal} {\bibinfo  {journal} {SciPost Phys.}\ }\textbf {\bibinfo {volume}
  {11}},\ \bibinfo {pages} {68} (\bibinfo {year} {2021})}\BibitemShut {NoStop}%
\bibitem [{\citenamefont {Pozsgay}\ \emph {et~al.}(2021)\citenamefont
  {Pozsgay}, \citenamefont {Gombor}, \citenamefont {Hutsalyuk}, \citenamefont
  {Jiang}, \citenamefont {Pristy\'ak},\ and\ \citenamefont
  {Vernier}}]{pozsgay2021integrable}%
  \BibitemOpen
  \bibfield  {author} {\bibinfo {author} {\bibfnamefont {B.}~\bibnamefont
  {Pozsgay}}, \bibinfo {author} {\bibfnamefont {T.}~\bibnamefont {Gombor}},
  \bibinfo {author} {\bibfnamefont {A.}~\bibnamefont {Hutsalyuk}}, \bibinfo
  {author} {\bibfnamefont {Y.}~\bibnamefont {Jiang}}, \bibinfo {author}
  {\bibfnamefont {L.}~\bibnamefont {Pristy\'ak}}, \ and\ \bibinfo {author}
  {\bibfnamefont {E.}~\bibnamefont {Vernier}},\ }\href {\doibase
  10.1103/PhysRevE.104.044106} {\bibfield  {journal} {\bibinfo  {journal}
  {Phys. Rev. E}\ }\textbf {\bibinfo {volume} {104}},\ \bibinfo {pages}
  {044106} (\bibinfo {year} {2021})}\BibitemShut {NoStop}%
\bibitem [{\citenamefont {Zadnik}\ and\ \citenamefont
  {Fagotti}(2021)}]{Zadnik2021folded1}%
  \BibitemOpen
  \bibfield  {author} {\bibinfo {author} {\bibfnamefont {L.}~\bibnamefont
  {Zadnik}}\ and\ \bibinfo {author} {\bibfnamefont {M.}~\bibnamefont
  {Fagotti}},\ }\href {\doibase 10.21468/SciPostPhysCore.4.2.010} {\bibfield
  {journal} {\bibinfo  {journal} {SciPost Phys. Core}\ }\textbf {\bibinfo
  {volume} {4}},\ \bibinfo {pages} {10} (\bibinfo {year} {2021})}\BibitemShut
  {NoStop}%
\bibitem [{\citenamefont {Zadnik}\ \emph {et~al.}(2021)\citenamefont {Zadnik},
  \citenamefont {Bidzhiev},\ and\ \citenamefont {Fagotti}}]{Zadnik2021folded2}%
  \BibitemOpen
  \bibfield  {author} {\bibinfo {author} {\bibfnamefont {L.}~\bibnamefont
  {Zadnik}}, \bibinfo {author} {\bibfnamefont {K.}~\bibnamefont {Bidzhiev}}, \
  and\ \bibinfo {author} {\bibfnamefont {M.}~\bibnamefont {Fagotti}},\ }\href
  {\doibase 10.21468/SciPostPhys.10.5.099} {\bibfield  {journal} {\bibinfo
  {journal} {SciPost Phys.}\ }\textbf {\bibinfo {volume} {10}},\ \bibinfo
  {pages} {99} (\bibinfo {year} {2021})}\BibitemShut {NoStop}%
\bibitem [{\citenamefont {Farrelly}(2020)}]{farrelly2020review}%
  \BibitemOpen
  \bibfield  {author} {\bibinfo {author} {\bibfnamefont {T.}~\bibnamefont
  {Farrelly}},\ }\href {\doibase 10.22331/q-2020-11-30-368} {\bibfield
  {journal} {\bibinfo  {journal} {Quantum}\ }\textbf {\bibinfo {volume} {4}},\
  \bibinfo {pages} {368} (\bibinfo {year} {2020})}\BibitemShut {NoStop}%
\bibitem [{\citenamefont {Piroli}\ and\ \citenamefont
  {Cirac}(2020)}]{piroli2020quantum}%
  \BibitemOpen
  \bibfield  {author} {\bibinfo {author} {\bibfnamefont {L.}~\bibnamefont
  {Piroli}}\ and\ \bibinfo {author} {\bibfnamefont {J.~I.}\ \bibnamefont
  {Cirac}},\ }\href {\doibase 10.1103/PhysRevLett.125.190402} {\bibfield
  {journal} {\bibinfo  {journal} {Phys. Rev. Lett.}\ }\textbf {\bibinfo
  {volume} {125}},\ \bibinfo {pages} {190402} (\bibinfo {year}
  {2020})}\BibitemShut {NoStop}%
\bibitem [{\citenamefont {Nahum}\ \emph {et~al.}(2017)\citenamefont {Nahum},
  \citenamefont {Ruhman}, \citenamefont {Vijay},\ and\ \citenamefont
  {Haah}}]{nahum2017quantum}%
  \BibitemOpen
  \bibfield  {author} {\bibinfo {author} {\bibfnamefont {A.}~\bibnamefont
  {Nahum}}, \bibinfo {author} {\bibfnamefont {J.}~\bibnamefont {Ruhman}},
  \bibinfo {author} {\bibfnamefont {S.}~\bibnamefont {Vijay}}, \ and\ \bibinfo
  {author} {\bibfnamefont {J.}~\bibnamefont {Haah}},\ }\href {\doibase
  10.1103/PhysRevX.7.031016} {\bibfield  {journal} {\bibinfo  {journal} {Phys.
  Rev. X}\ }\textbf {\bibinfo {volume} {7}},\ \bibinfo {pages} {031016}
  (\bibinfo {year} {2017})}\BibitemShut {NoStop}%
\bibitem [{\citenamefont {Nahum}\ \emph {et~al.}(2018)\citenamefont {Nahum},
  \citenamefont {Vijay},\ and\ \citenamefont {Haah}}]{nahum2018operator}%
  \BibitemOpen
  \bibfield  {author} {\bibinfo {author} {\bibfnamefont {A.}~\bibnamefont
  {Nahum}}, \bibinfo {author} {\bibfnamefont {S.}~\bibnamefont {Vijay}}, \ and\
  \bibinfo {author} {\bibfnamefont {J.}~\bibnamefont {Haah}},\ }\href {\doibase
  10.1103/PhysRevX.8.021014} {\bibfield  {journal} {\bibinfo  {journal} {Phys.
  Rev. X}\ }\textbf {\bibinfo {volume} {8}},\ \bibinfo {pages} {021014}
  (\bibinfo {year} {2018})}\BibitemShut {NoStop}%
\bibitem [{\citenamefont {von Keyserlingk}\ \emph {et~al.}(2018)\citenamefont
  {von Keyserlingk}, \citenamefont {Rakovszky}, \citenamefont {Pollmann},\ and\
  \citenamefont {Sondhi}}]{vonkeyserlingk2018operator}%
  \BibitemOpen
  \bibfield  {author} {\bibinfo {author} {\bibfnamefont {C.~W.}\ \bibnamefont
  {von Keyserlingk}}, \bibinfo {author} {\bibfnamefont {T.}~\bibnamefont
  {Rakovszky}}, \bibinfo {author} {\bibfnamefont {F.}~\bibnamefont {Pollmann}},
  \ and\ \bibinfo {author} {\bibfnamefont {S.~L.}\ \bibnamefont {Sondhi}},\
  }\href {\doibase 10.1103/PhysRevX.8.021013} {\bibfield  {journal} {\bibinfo
  {journal} {Phys. Rev. X}\ }\textbf {\bibinfo {volume} {8}},\ \bibinfo {pages}
  {021013} (\bibinfo {year} {2018})}\BibitemShut {NoStop}%
\bibitem [{\citenamefont {Bertini}\ \emph
  {et~al.}(2019{\natexlab{a}})\citenamefont {Bertini}, \citenamefont {Kos},\
  and\ \citenamefont {Prosen}}]{bertini2019exact}%
  \BibitemOpen
  \bibfield  {author} {\bibinfo {author} {\bibfnamefont {B.}~\bibnamefont
  {Bertini}}, \bibinfo {author} {\bibfnamefont {P.}~\bibnamefont {Kos}}, \ and\
  \bibinfo {author} {\bibfnamefont {T.}~\bibnamefont {Prosen}},\ }\href
  {\doibase 10.1103/PhysRevLett.123.210601} {\bibfield  {journal} {\bibinfo
  {journal} {Phys. Rev. Lett.}\ }\textbf {\bibinfo {volume} {123}},\ \bibinfo
  {pages} {210601} (\bibinfo {year} {2019}{\natexlab{a}})}\BibitemShut
  {NoStop}%
\bibitem [{\citenamefont {Bertini}\ \emph
  {et~al.}(2019{\natexlab{b}})\citenamefont {Bertini}, \citenamefont {Kos},\
  and\ \citenamefont {Prosen}}]{bertini2019entanglement}%
  \BibitemOpen
  \bibfield  {author} {\bibinfo {author} {\bibfnamefont {B.}~\bibnamefont
  {Bertini}}, \bibinfo {author} {\bibfnamefont {P.}~\bibnamefont {Kos}}, \ and\
  \bibinfo {author} {\bibfnamefont {T.}~\bibnamefont {Prosen}},\ }\href
  {\doibase 10.1103/PhysRevX.9.021033} {\bibfield  {journal} {\bibinfo
  {journal} {Phys. Rev. X}\ }\textbf {\bibinfo {volume} {9}},\ \bibinfo {pages}
  {021033} (\bibinfo {year} {2019}{\natexlab{b}})}\BibitemShut {NoStop}%
\bibitem [{\citenamefont {Piroli}\ \emph {et~al.}(2020)\citenamefont {Piroli},
  \citenamefont {Bertini}, \citenamefont {Cirac},\ and\ \citenamefont
  {Prosen}}]{piroli2020exact}%
  \BibitemOpen
  \bibfield  {author} {\bibinfo {author} {\bibfnamefont {L.}~\bibnamefont
  {Piroli}}, \bibinfo {author} {\bibfnamefont {B.}~\bibnamefont {Bertini}},
  \bibinfo {author} {\bibfnamefont {J.~I.}\ \bibnamefont {Cirac}}, \ and\
  \bibinfo {author} {\bibfnamefont {T.}~\bibnamefont {Prosen}},\ }\href
  {\doibase 10.1103/PhysRevB.101.094304} {\bibfield  {journal} {\bibinfo
  {journal} {Phys. Rev. B}\ }\textbf {\bibinfo {volume} {101}},\ \bibinfo
  {pages} {094304} (\bibinfo {year} {2020})}\BibitemShut {NoStop}%
\bibitem [{\citenamefont {Kos}\ \emph {et~al.}(2021)\citenamefont {Kos},
  \citenamefont {Bertini},\ and\ \citenamefont
  {Prosen}}]{pavel2021correlations}%
  \BibitemOpen
  \bibfield  {author} {\bibinfo {author} {\bibfnamefont {P.}~\bibnamefont
  {Kos}}, \bibinfo {author} {\bibfnamefont {B.}~\bibnamefont {Bertini}}, \ and\
  \bibinfo {author} {\bibfnamefont {T.}~\bibnamefont {Prosen}},\ }\href
  {\doibase 10.1103/PhysRevX.11.011022} {\bibfield  {journal} {\bibinfo
  {journal} {Phys. Rev. X}\ }\textbf {\bibinfo {volume} {11}},\ \bibinfo
  {pages} {011022} (\bibinfo {year} {2021})}\BibitemShut {NoStop}%
\bibitem [{\citenamefont {Claeys}\ and\ \citenamefont
  {Lamacraft}(2021)}]{claeys2021ergodic}%
  \BibitemOpen
  \bibfield  {author} {\bibinfo {author} {\bibfnamefont {P.~W.}\ \bibnamefont
  {Claeys}}\ and\ \bibinfo {author} {\bibfnamefont {A.}~\bibnamefont
  {Lamacraft}},\ }\href {\doibase 10.1103/PhysRevLett.126.100603} {\bibfield
  {journal} {\bibinfo  {journal} {Phys. Rev. Lett.}\ }\textbf {\bibinfo
  {volume} {126}},\ \bibinfo {pages} {100603} (\bibinfo {year}
  {2021})}\BibitemShut {NoStop}%
\bibitem [{\citenamefont {Friedman}\ \emph {et~al.}(2019)\citenamefont
  {Friedman}, \citenamefont {Gopalakrishnan},\ and\ \citenamefont
  {Vasseur}}]{friedman2019integrable}%
  \BibitemOpen
  \bibfield  {author} {\bibinfo {author} {\bibfnamefont {A.~J.}\ \bibnamefont
  {Friedman}}, \bibinfo {author} {\bibfnamefont {S.}~\bibnamefont
  {Gopalakrishnan}}, \ and\ \bibinfo {author} {\bibfnamefont {R.}~\bibnamefont
  {Vasseur}},\ }\href {\doibase 10.1103/PhysRevLett.123.170603} {\bibfield
  {journal} {\bibinfo  {journal} {Phys. Rev. Lett.}\ }\textbf {\bibinfo
  {volume} {123}},\ \bibinfo {pages} {170603} (\bibinfo {year}
  {2019})}\BibitemShut {NoStop}%
\bibitem [{\citenamefont {Klobas}\ \emph {et~al.}(2021)\citenamefont {Klobas},
  \citenamefont {Bertini},\ and\ \citenamefont {Piroli}}]{klobas2021exact}%
  \BibitemOpen
  \bibfield  {author} {\bibinfo {author} {\bibfnamefont {K.}~\bibnamefont
  {Klobas}}, \bibinfo {author} {\bibfnamefont {B.}~\bibnamefont {Bertini}}, \
  and\ \bibinfo {author} {\bibfnamefont {L.}~\bibnamefont {Piroli}},\ }\href
  {\doibase 10.1103/PhysRevLett.126.160602} {\bibfield  {journal} {\bibinfo
  {journal} {Phys. Rev. Lett.}\ }\textbf {\bibinfo {volume} {126}},\ \bibinfo
  {pages} {160602} (\bibinfo {year} {2021})}\BibitemShut {NoStop}%
\bibitem [{\citenamefont {Klobas}\ and\ \citenamefont
  {Bertini}(2021{\natexlab{a}})}]{klobas2021exact2}%
  \BibitemOpen
  \bibfield  {author} {\bibinfo {author} {\bibfnamefont {K.}~\bibnamefont
  {Klobas}}\ and\ \bibinfo {author} {\bibfnamefont {B.}~\bibnamefont
  {Bertini}},\ }\href {\doibase 10.21468/SciPostPhys.11.6.106} {\bibfield
  {journal} {\bibinfo  {journal} {SciPost Phys.}\ }\textbf {\bibinfo {volume}
  {11}},\ \bibinfo {pages} {106} (\bibinfo {year}
  {2021}{\natexlab{a}})}\BibitemShut {NoStop}%
\bibitem [{\citenamefont {Klobas}\ and\ \citenamefont
  {Bertini}(2021{\natexlab{b}})}]{klobas2021ent}%
  \BibitemOpen
  \bibfield  {author} {\bibinfo {author} {\bibfnamefont {K.}~\bibnamefont
  {Klobas}}\ and\ \bibinfo {author} {\bibfnamefont {B.}~\bibnamefont
  {Bertini}},\ }\href {\doibase 10.21468/SciPostPhys.11.6.107} {\bibfield
  {journal} {\bibinfo  {journal} {SciPost Phys.}\ }\textbf {\bibinfo {volume}
  {11}},\ \bibinfo {pages} {107} (\bibinfo {year}
  {2021}{\natexlab{b}})}\BibitemShut {NoStop}%
\bibitem [{\citenamefont {Gombor}\ and\ \citenamefont
  {Pozsgay}(2021)}]{gombor2021integrable}%
  \BibitemOpen
  \bibfield  {author} {\bibinfo {author} {\bibfnamefont {T.}~\bibnamefont
  {Gombor}}\ and\ \bibinfo {author} {\bibfnamefont {B.}~\bibnamefont
  {Pozsgay}},\ }\href {\doibase 10.1103/PhysRevE.104.054123} {\bibfield
  {journal} {\bibinfo  {journal} {Phys. Rev. E}\ }\textbf {\bibinfo {volume}
  {104}},\ \bibinfo {pages} {054123} (\bibinfo {year} {2021})}\BibitemShut
  {NoStop}%
\bibitem [{\citenamefont {Prosen}(2021)}]{prosen2021reversible}%
  \BibitemOpen
  \bibfield  {author} {\bibinfo {author} {\bibfnamefont {T.}~\bibnamefont
  {Prosen}},\ }\href {https://arxiv.org/abs/2106.01292} {\bibfield  {journal}
  {\bibinfo  {journal} {arXiv:2106.01292}\ } (\bibinfo {year}
  {2021})}\BibitemShut {NoStop}%
\bibitem [{\citenamefont {Gombor}\ and\ \citenamefont
  {Pozsgay}(2022)}]{gombor2022integrable}%
  \BibitemOpen
  \bibfield  {author} {\bibinfo {author} {\bibfnamefont {T.}~\bibnamefont
  {Gombor}}\ and\ \bibinfo {author} {\bibfnamefont {B.}~\bibnamefont
  {Pozsgay}},\ }\href {https://arxiv.org/abs/2205.02038} {\bibfield  {journal}
  {\bibinfo  {journal} {arXiv:2205.02038}\ } (\bibinfo {year}
  {2022})}\BibitemShut {NoStop}%
\bibitem [{\citenamefont {Brennen}\ and\ \citenamefont
  {Williams}(2003)}]{brennen2003entanglement}%
  \BibitemOpen
  \bibfield  {author} {\bibinfo {author} {\bibfnamefont {G.~K.}\ \bibnamefont
  {Brennen}}\ and\ \bibinfo {author} {\bibfnamefont {J.~E.}\ \bibnamefont
  {Williams}},\ }\href {\doibase 10.1103/PhysRevA.68.042311} {\bibfield
  {journal} {\bibinfo  {journal} {Phys. Rev. A}\ }\textbf {\bibinfo {volume}
  {68}},\ \bibinfo {pages} {042311} (\bibinfo {year} {2003})}\BibitemShut
  {NoStop}%
\bibitem [{\citenamefont {Li}\ \emph {et~al.}(2018)\citenamefont {Li},
  \citenamefont {Chen},\ and\ \citenamefont {Fisher}}]{li2018quantum}%
  \BibitemOpen
  \bibfield  {author} {\bibinfo {author} {\bibfnamefont {Y.}~\bibnamefont
  {Li}}, \bibinfo {author} {\bibfnamefont {X.}~\bibnamefont {Chen}}, \ and\
  \bibinfo {author} {\bibfnamefont {M.~P.~A.}\ \bibnamefont {Fisher}},\ }\href
  {\doibase 10.1103/PhysRevB.98.205136} {\bibfield  {journal} {\bibinfo
  {journal} {Phys. Rev. B}\ }\textbf {\bibinfo {volume} {98}},\ \bibinfo
  {pages} {205136} (\bibinfo {year} {2018})}\BibitemShut {NoStop}%
\bibitem [{\citenamefont {Skinner}\ \emph {et~al.}(2019)\citenamefont
  {Skinner}, \citenamefont {Ruhman},\ and\ \citenamefont
  {Nahum}}]{skinner2019meas}%
  \BibitemOpen
  \bibfield  {author} {\bibinfo {author} {\bibfnamefont {B.}~\bibnamefont
  {Skinner}}, \bibinfo {author} {\bibfnamefont {J.}~\bibnamefont {Ruhman}}, \
  and\ \bibinfo {author} {\bibfnamefont {A.}~\bibnamefont {Nahum}},\ }\href
  {\doibase 10.1103/PhysRevX.9.031009} {\bibfield  {journal} {\bibinfo
  {journal} {Phys. Rev. X}\ }\textbf {\bibinfo {volume} {9}},\ \bibinfo {pages}
  {031009} (\bibinfo {year} {2019})}\BibitemShut {NoStop}%
\bibitem [{\citenamefont {Chan}\ \emph {et~al.}(2019)\citenamefont {Chan},
  \citenamefont {Nandkishore}, \citenamefont {Pretko},\ and\ \citenamefont
  {Smith}}]{chan2019unitary}%
  \BibitemOpen
  \bibfield  {author} {\bibinfo {author} {\bibfnamefont {A.}~\bibnamefont
  {Chan}}, \bibinfo {author} {\bibfnamefont {R.~M.}\ \bibnamefont
  {Nandkishore}}, \bibinfo {author} {\bibfnamefont {M.}~\bibnamefont {Pretko}},
  \ and\ \bibinfo {author} {\bibfnamefont {G.}~\bibnamefont {Smith}},\ }\href
  {\doibase 10.1103/PhysRevB.99.224307} {\bibfield  {journal} {\bibinfo
  {journal} {Phys. Rev. B}\ }\textbf {\bibinfo {volume} {99}},\ \bibinfo
  {pages} {224307} (\bibinfo {year} {2019})}\BibitemShut {NoStop}%
\bibitem [{\citenamefont {Jian}\ \emph {et~al.}(2020)\citenamefont {Jian},
  \citenamefont {Bauer}, \citenamefont {Keselman},\ and\ \citenamefont
  {Ludwig}}]{jian2020criticality}%
  \BibitemOpen
  \bibfield  {author} {\bibinfo {author} {\bibfnamefont {C.-M.}\ \bibnamefont
  {Jian}}, \bibinfo {author} {\bibfnamefont {B.}~\bibnamefont {Bauer}},
  \bibinfo {author} {\bibfnamefont {A.}~\bibnamefont {Keselman}}, \ and\
  \bibinfo {author} {\bibfnamefont {A.~W.}\ \bibnamefont {Ludwig}},\ }\href
  {https://arxiv.org/abs/2012.04666} {\bibfield  {journal} {\bibinfo  {journal}
  {arXiv:2012.04666}\ } (\bibinfo {year} {2020})}\BibitemShut {NoStop}%
\bibitem [{\citenamefont {Ippoliti}\ and\ \citenamefont
  {Khemani}(2021)}]{Ippoliti2021post}%
  \BibitemOpen
  \bibfield  {author} {\bibinfo {author} {\bibfnamefont {M.}~\bibnamefont
  {Ippoliti}}\ and\ \bibinfo {author} {\bibfnamefont {V.}~\bibnamefont
  {Khemani}},\ }\href {\doibase 10.1103/PhysRevLett.126.060501} {\bibfield
  {journal} {\bibinfo  {journal} {Phys. Rev. Lett.}\ }\textbf {\bibinfo
  {volume} {126}},\ \bibinfo {pages} {060501} (\bibinfo {year}
  {2021})}\BibitemShut {NoStop}%
\bibitem [{\citenamefont {Lu}\ and\ \citenamefont
  {Grover}(2021)}]{lu2021spacetime}%
  \BibitemOpen
  \bibfield  {author} {\bibinfo {author} {\bibfnamefont {T.-C.}\ \bibnamefont
  {Lu}}\ and\ \bibinfo {author} {\bibfnamefont {T.}~\bibnamefont {Grover}},\
  }\href {\doibase 10.1103/PRXQuantum.2.040319} {\bibfield  {journal} {\bibinfo
   {journal} {PRX Quantum}\ }\textbf {\bibinfo {volume} {2}},\ \bibinfo {pages}
  {040319} (\bibinfo {year} {2021})}\BibitemShut {NoStop}%
\bibitem [{\citenamefont {Ippoliti}\ \emph {et~al.}(2022)\citenamefont
  {Ippoliti}, \citenamefont {Rakovszky},\ and\ \citenamefont
  {Khemani}}]{Ippoliti2022fractal}%
  \BibitemOpen
  \bibfield  {author} {\bibinfo {author} {\bibfnamefont {M.}~\bibnamefont
  {Ippoliti}}, \bibinfo {author} {\bibfnamefont {T.}~\bibnamefont {Rakovszky}},
  \ and\ \bibinfo {author} {\bibfnamefont {V.}~\bibnamefont {Khemani}},\ }\href
  {\doibase 10.1103/PhysRevX.12.011045} {\bibfield  {journal} {\bibinfo
  {journal} {Phys. Rev. X}\ }\textbf {\bibinfo {volume} {12}},\ \bibinfo
  {pages} {011045} (\bibinfo {year} {2022})}\BibitemShut {NoStop}%
\bibitem [{\citenamefont {Medvedyeva}\ \emph {et~al.}(2016)\citenamefont
  {Medvedyeva}, \citenamefont {Essler},\ and\ \citenamefont
  {Prosen}}]{Medvedyeva2016exact}%
  \BibitemOpen
  \bibfield  {author} {\bibinfo {author} {\bibfnamefont {M.~V.}\ \bibnamefont
  {Medvedyeva}}, \bibinfo {author} {\bibfnamefont {F.~H.~L.}\ \bibnamefont
  {Essler}}, \ and\ \bibinfo {author} {\bibfnamefont {T.~c.~v.}\ \bibnamefont
  {Prosen}},\ }\href {\doibase 10.1103/PhysRevLett.117.137202} {\bibfield
  {journal} {\bibinfo  {journal} {Phys. Rev. Lett.}\ }\textbf {\bibinfo
  {volume} {117}},\ \bibinfo {pages} {137202} (\bibinfo {year}
  {2016})}\BibitemShut {NoStop}%
\bibitem [{\citenamefont {S\'a}\ \emph {et~al.}(2021)\citenamefont {S\'a},
  \citenamefont {Ribeiro},\ and\ \citenamefont {Prosen}}]{sa2021integrable}%
  \BibitemOpen
  \bibfield  {author} {\bibinfo {author} {\bibfnamefont {L.}~\bibnamefont
  {S\'a}}, \bibinfo {author} {\bibfnamefont {P.}~\bibnamefont {Ribeiro}}, \
  and\ \bibinfo {author} {\bibfnamefont {T.}~\bibnamefont {Prosen}},\ }\href
  {\doibase 10.1103/PhysRevB.103.115132} {\bibfield  {journal} {\bibinfo
  {journal} {Phys. Rev. B}\ }\textbf {\bibinfo {volume} {103}},\ \bibinfo
  {pages} {115132} (\bibinfo {year} {2021})}\BibitemShut {NoStop}%
\bibitem [{\citenamefont {Ziolkowska}\ and\ \citenamefont
  {Essler}(2020)}]{ziolkowska2020yang}%
  \BibitemOpen
  \bibfield  {author} {\bibinfo {author} {\bibfnamefont {A.~A.}\ \bibnamefont
  {Ziolkowska}}\ and\ \bibinfo {author} {\bibfnamefont {F.}~\bibnamefont
  {Essler}},\ }\href {\doibase 10.21468/SciPostPhys.8.3.044} {\bibfield
  {journal} {\bibinfo  {journal} {SciPost}\ }\textbf {\bibinfo {volume} {8}},\
  \bibinfo {pages} {044} (\bibinfo {year} {2020})}\BibitemShut {NoStop}%
\bibitem [{\citenamefont {de~Leeuw}\ \emph {et~al.}(2021)\citenamefont
  {de~Leeuw}, \citenamefont {Paletta},\ and\ \citenamefont
  {Pozsgay}}]{de2021constructing}%
  \BibitemOpen
  \bibfield  {author} {\bibinfo {author} {\bibfnamefont {M.}~\bibnamefont
  {de~Leeuw}}, \bibinfo {author} {\bibfnamefont {C.}~\bibnamefont {Paletta}}, \
  and\ \bibinfo {author} {\bibfnamefont {B.}~\bibnamefont {Pozsgay}},\ }\href
  {\doibase 10.1103/PhysRevLett.126.240403} {\bibfield  {journal} {\bibinfo
  {journal} {Phys. Rev. Lett.}\ }\textbf {\bibinfo {volume} {126}},\ \bibinfo
  {pages} {240403} (\bibinfo {year} {2021})}\BibitemShut {NoStop}%
\bibitem [{\citenamefont {Bobenko}\ \emph {et~al.}(1993)\citenamefont
  {Bobenko}, \citenamefont {Bordemann}, \citenamefont {Gunn},\ and\
  \citenamefont {Pinkall}}]{bobenko1993two}%
  \BibitemOpen
  \bibfield  {author} {\bibinfo {author} {\bibfnamefont {A.}~\bibnamefont
  {Bobenko}}, \bibinfo {author} {\bibfnamefont {M.}~\bibnamefont {Bordemann}},
  \bibinfo {author} {\bibfnamefont {C.}~\bibnamefont {Gunn}}, \ and\ \bibinfo
  {author} {\bibfnamefont {U.}~\bibnamefont {Pinkall}},\ }\href {\doibase
  10.1007/BF02097234} {\bibfield  {journal} {\bibinfo  {journal} {Commun. Math.
  Phys.}\ }\textbf {\bibinfo {volume} {158}},\ \bibinfo {pages} {127} (\bibinfo
  {year} {1993})}\BibitemShut {NoStop}%
\bibitem [{\citenamefont {Vanicat}\ \emph {et~al.}(2018)\citenamefont
  {Vanicat}, \citenamefont {Zadnik},\ and\ \citenamefont
  {Prosen}}]{vanicat2018integrable}%
  \BibitemOpen
  \bibfield  {author} {\bibinfo {author} {\bibfnamefont {M.}~\bibnamefont
  {Vanicat}}, \bibinfo {author} {\bibfnamefont {L.}~\bibnamefont {Zadnik}}, \
  and\ \bibinfo {author} {\bibfnamefont {T.}~\bibnamefont {Prosen}},\ }\href
  {\doibase 10.1103/PhysRevLett.121.030606} {\bibfield  {journal} {\bibinfo
  {journal} {Phys. Rev. Lett.}\ }\textbf {\bibinfo {volume} {121}},\ \bibinfo
  {pages} {030606} (\bibinfo {year} {2018})}\BibitemShut {NoStop}%
\bibitem [{\citenamefont {de~Leeuw}\ \emph {et~al.}(2019)\citenamefont
  {de~Leeuw}, \citenamefont {Pribytok},\ and\ \citenamefont
  {Ryan}}]{de2019classifying}%
  \BibitemOpen
  \bibfield  {author} {\bibinfo {author} {\bibfnamefont {M.}~\bibnamefont
  {de~Leeuw}}, \bibinfo {author} {\bibfnamefont {A.}~\bibnamefont {Pribytok}},
  \ and\ \bibinfo {author} {\bibfnamefont {P.}~\bibnamefont {Ryan}},\ }\href
  {\doibase 10.1088/1751-8121/ab529f} {\bibfield  {journal} {\bibinfo
  {journal} {J. Phys. A}\ }\textbf {\bibinfo {volume} {52}},\ \bibinfo {pages}
  {505201} (\bibinfo {year} {2019})}\BibitemShut {NoStop}%
\bibitem [{\citenamefont {de~Leeuw}\ \emph {et~al.}(2020)\citenamefont
  {de~Leeuw}, \citenamefont {Paletta}, \citenamefont {Pribytok}, \citenamefont
  {Retore},\ and\ \citenamefont {Ryan}}]{de2020classifying}%
  \BibitemOpen
  \bibfield  {author} {\bibinfo {author} {\bibfnamefont {M.}~\bibnamefont
  {de~Leeuw}}, \bibinfo {author} {\bibfnamefont {C.}~\bibnamefont {Paletta}},
  \bibinfo {author} {\bibfnamefont {A.}~\bibnamefont {Pribytok}}, \bibinfo
  {author} {\bibfnamefont {A.~L.}\ \bibnamefont {Retore}}, \ and\ \bibinfo
  {author} {\bibfnamefont {P.}~\bibnamefont {Ryan}},\ }\href {\doibase
  10.1103/PhysRevLett.125.031604} {\bibfield  {journal} {\bibinfo  {journal}
  {Phys. Rev. Lett.}\ }\textbf {\bibinfo {volume} {125}},\ \bibinfo {pages}
  {031604} (\bibinfo {year} {2020})}\BibitemShut {NoStop}%
\bibitem [{\citenamefont {Preskill}(1998)}]{preskill1998lecture}%
  \BibitemOpen
  \bibfield  {author} {\bibinfo {author} {\bibfnamefont {J.}~\bibnamefont
  {Preskill}},\ }\href@noop {} {\emph {\bibinfo {title} {Lecture notes for
  physics 229: Quantum information and computation}}}\ (\bibinfo {year}
  {1998})\BibitemShut {NoStop}%
\bibitem [{Note1()}]{Note1}%
  \BibitemOpen
  \bibinfo {note} {Note that if $\protect \check {R}$ has a difference form,
  i.e., $\protect \check {R}(u, v) = \protect \check {R}(u-v)$, we simply write
  $\protect \check {R}(u) = \protect \check {R}(u, 0)$. Similarly for $K_a$ and
  other operators. In this case, ${\protect \cal {L}}$ is independent on
  $u$.}\BibitemShut {Stop}%
\bibitem [{\citenamefont {Mendl}\ and\ \citenamefont
  {Wolf}(2009)}]{mendl2009unital}%
  \BibitemOpen
  \bibfield  {author} {\bibinfo {author} {\bibfnamefont {C.~B.}\ \bibnamefont
  {Mendl}}\ and\ \bibinfo {author} {\bibfnamefont {M.~M.}\ \bibnamefont
  {Wolf}},\ }\href {\doibase 10.1007/s00220-009-0824-2} {\bibfield  {journal}
  {\bibinfo  {journal} {Commun. Math. Phys.}\ }\textbf {\bibinfo {volume}
  {289}},\ \bibinfo {pages} {1057} (\bibinfo {year} {2009})}\BibitemShut
  {NoStop}%
\bibitem [{Note2()}]{Note2}%
  \BibitemOpen
  \bibinfo {note} {There are typos for the $R$-matrix of Model B3 in Ref. \cite
  {de2021constructing} : $R_4^{10} = e^{\protect \text {i}\phi } 2\protect
  \text {i}(\gamma +1)^2 \protect \qopname \relax o{sinh}(\alpha )/\zeta $ and
  $R_4^7 = -e^{-\protect \text {i}\phi } 2\protect \text {i}(\gamma +1)^2
  \protect \qopname \relax o{sinh}(\alpha )/\zeta $.}\BibitemShut {Stop}%
\bibitem [{\citenamefont {Fradkin}(2013)}]{fradkin2013field}%
  \BibitemOpen
  \bibfield  {author} {\bibinfo {author} {\bibfnamefont {E.}~\bibnamefont
  {Fradkin}},\ }\href@noop {} {\emph {\bibinfo {title} {Field theories of
  condensed matter physics}}}\ (\bibinfo  {publisher} {Cambridge University
  Press},\ \bibinfo {year} {2013})\BibitemShut {NoStop}%
\bibitem [{\citenamefont {Wolfram}(1983)}]{Wolfram}%
  \BibitemOpen
  \bibfield  {author} {\bibinfo {author} {\bibfnamefont {S.}~\bibnamefont
  {Wolfram}},\ }\href {\doibase 10.1103/RevModPhys.55.601} {\bibfield
  {journal} {\bibinfo  {journal} {Rev. Mod. Phys.}\ }\textbf {\bibinfo {volume}
  {55}},\ \bibinfo {pages} {601} (\bibinfo {year} {1983})}\BibitemShut
  {NoStop}%
\bibitem [{\citenamefont {Wilkinson}\ \emph {et~al.}(2021)\citenamefont
  {Wilkinson}, \citenamefont {Prosen},\ and\ \citenamefont
  {Garrahan}}]{wilkinson2021exact}%
  \BibitemOpen
  \bibfield  {author} {\bibinfo {author} {\bibfnamefont {J.~W.}\ \bibnamefont
  {Wilkinson}}, \bibinfo {author} {\bibfnamefont {T.}~\bibnamefont {Prosen}}, \
  and\ \bibinfo {author} {\bibfnamefont {J.~P.}\ \bibnamefont {Garrahan}},\
  }\href {https://arxiv.org/abs/2110.15085} {\bibfield  {journal} {\bibinfo
  {journal} {arXiv:2110.15085}\ } (\bibinfo {year} {2021})}\BibitemShut
  {NoStop}%
\bibitem [{\citenamefont {Jones}\ and\ \citenamefont
  {Linden}(2021)}]{jones2021integrable}%
  \BibitemOpen
  \bibfield  {author} {\bibinfo {author} {\bibfnamefont {N.~G.}\ \bibnamefont
  {Jones}}\ and\ \bibinfo {author} {\bibfnamefont {N.}~\bibnamefont {Linden}},\
  }\href {https://arxiv.org/abs/2107.02184} {\bibfield  {journal} {\bibinfo
  {journal} {arXiv:2107.02184}\ } (\bibinfo {year} {2021})}\BibitemShut
  {NoStop}%
\bibitem [{\citenamefont {Essler}\ and\ \citenamefont
  {Piroli}(2020)}]{Essler2020integrability}%
  \BibitemOpen
  \bibfield  {author} {\bibinfo {author} {\bibfnamefont {F.~H.~L.}\
  \bibnamefont {Essler}}\ and\ \bibinfo {author} {\bibfnamefont
  {L.}~\bibnamefont {Piroli}},\ }\href {\doibase 10.1103/PhysRevE.102.062210}
  {\bibfield  {journal} {\bibinfo  {journal} {Phys. Rev. E}\ }\textbf {\bibinfo
  {volume} {102}},\ \bibinfo {pages} {062210} (\bibinfo {year}
  {2020})}\BibitemShut {NoStop}%
\bibitem [{\citenamefont {Wiseman}(1996)}]{wiseman1996quantum}%
  \BibitemOpen
  \bibfield  {author} {\bibinfo {author} {\bibfnamefont {H.~M.}\ \bibnamefont
  {Wiseman}},\ }\href
  {https://iopscience.iop.org/article/10.1088/1355-5111/8/1/015/meta?casa_token=rFb_zqpLp1AAAAAA:4diQT8O6D5c7xhHhbhT8ZbttDiU0FlpZAvKcQHhN2UgPWAb_gXzUmF5KAdvTSgokHLpU8LP0NC1PmJb-6AA}
  {\bibfield  {journal} {\bibinfo  {journal} {Quantum Semiclass. Opt.}\
  }\textbf {\bibinfo {volume} {8}},\ \bibinfo {pages} {205} (\bibinfo {year}
  {1996})}\BibitemShut {NoStop}%
\bibitem [{\citenamefont {Claeys}\ \emph {et~al.}(2022)\citenamefont {Claeys},
  \citenamefont {Herzog-Arbeitman},\ and\ \citenamefont
  {Lamacraft}}]{claeys2022correlations}%
  \BibitemOpen
  \bibfield  {author} {\bibinfo {author} {\bibfnamefont {P.~W.}\ \bibnamefont
  {Claeys}}, \bibinfo {author} {\bibfnamefont {J.}~\bibnamefont
  {Herzog-Arbeitman}}, \ and\ \bibinfo {author} {\bibfnamefont
  {A.}~\bibnamefont {Lamacraft}},\ }\href {\doibase
  10.21468/SciPostPhys.12.1.007} {\bibfield  {journal} {\bibinfo  {journal}
  {SciPost Phys.}\ }\textbf {\bibinfo {volume} {12}},\ \bibinfo {pages} {7}
  (\bibinfo {year} {2022})}\BibitemShut {NoStop}%
\bibitem [{\citenamefont {Causer}\ \emph {et~al.}(2020)\citenamefont {Causer},
  \citenamefont {Lesanovsky}, \citenamefont {Ba\~nuls},\ and\ \citenamefont
  {Garrahan}}]{causer2020dynamics}%
  \BibitemOpen
  \bibfield  {author} {\bibinfo {author} {\bibfnamefont {L.}~\bibnamefont
  {Causer}}, \bibinfo {author} {\bibfnamefont {I.}~\bibnamefont {Lesanovsky}},
  \bibinfo {author} {\bibfnamefont {M.~C.}\ \bibnamefont {Ba\~nuls}}, \ and\
  \bibinfo {author} {\bibfnamefont {J.~P.}\ \bibnamefont {Garrahan}},\ }\href
  {\doibase 10.1103/PhysRevE.102.052132} {\bibfield  {journal} {\bibinfo
  {journal} {Phys. Rev. E}\ }\textbf {\bibinfo {volume} {102}},\ \bibinfo
  {pages} {052132} (\bibinfo {year} {2020})}\BibitemShut {NoStop}%
\end{thebibliography}%


%

\end{document}